\documentclass{aa}  
\usepackage[varg]{txfonts}
\usepackage{graphicx}
\usepackage{txfonts}
\usepackage{float}
\usepackage[colorlinks=true,allcolors=blue]{hyperref}
\usepackage{multirow}
\usepackage{color}
\usepackage{csquotes}
\usepackage{caption}
\usepackage{comment}
\usepackage{longtable}

\begin{document}

\title{Activity correlation and temporal variation of small-scale magnetic fields on young Sun-like stars}
\subtitle{}

   \author{
   A. Hahlin\inst{1,2}
   \and B. Zaire\inst{3}
   \and C. P. Folsom\inst{4}
   \and K. Al Moulla\inst{5}\thanks{SNSF Postdoctoral Fellow}
   \and A. Lavail\inst{6}
   }
   \institute{
   Department of Physics and Astronomy, Uppsala University, Box 516, SE-751 20 Uppsala, Sweden 
   \and Astrophysics Group, Keele University, Staffordshire ST5 5BG, UK\\\email{a.j.hahlin@keele.ac.uk}
   \and Departamento de Física, Universidade Federal de Minas Gerais, 31270-901 Belo Horizonte, Minas Gerais, Brazil
   \and Tartu Observatory, University of Tartu, Observatooriumi 1, 61602, Toravere, Estonia
   \and Instituto de Astrofísica e Ciências do Espaço, Universidade do Porto, CAUP, Rua das Estrelas, 4150-762 Porto, Portugal
   \and Namzitu astro, 31130 Quint-Fonsegrives, France
   \\
   }

    \date{xxxx-xx-xx}
    
\abstract
{}
{We aim to evaluate how well the variation of small-scale magnetic fields on the stellar surface can be monitored with time-series observations. Further, we aim to establish to what extent the measured total unsigned magnetic field traces other activity indicators.}
{We measured the total unsigned magnetic field on four young, Sun-like, stars using the Zeeman splitting of magnetically sensitive \ion{Ti}{I} and \ion{Fe}{I} lines from high-resolution time series spectra obtained with the spectropolarimeters ESPaDOnS at Canada France Hawaii Telescope and NARVAL at Bernard Lyot Telescope. We then characterised the magnetic field variations using both sinusoidal variation and Lomb-Scargle periodograms. We evaluated how the rotational variation of the {total unsigned} magnetic field strength correlates with the activity indicators S-index, H$\alpha$-index, Ca IRT-index, and the large-scale magnetic field obtained from Zeeman Doppler imaging maps obtained in earlier studies.}
{We find clear signals of rotational modulation of the total magnetic field on HIP 76768 and tentative detection on Mel 25-5. This is supported both by the sinusoidal fitting as well as the periodogram. For the other stars, we find no clear modulation signals of the total magnetic field. We find positive correlations between the total magnetic field and activity indices on all four stars, indicating that indirect magnetic activity indicators trace the underlying magnetic field variability. 
However, comparing the activity-magnetic field relationship between the stars in our sample shows a significant deviation between activity level and measured magnetic field strength.}
{Small-scale magnetic field variability can evidently be traced using the Zeeman effect on magnetically sensitive lines, provided that the star is sufficiently active. It is also possible to self-consistently recover rotational periods from such measurements. The primary limit for the detection of magnetic field variations on less active stars is the precision of Zeeman broadening and intensification measurements.} 


\authorrunning{A. Hahlin et al.}
\maketitle

\section{Introduction}

Signals of magnetic activity on stellar surfaces are known to vary over different timescales. This variability can be measured in several ways, such as starspots \citep[e.g.][]{Messina:2002}, emission in chromospheric lines \citep[e.g.][]{Baliunas:1995}, and even from direct measurements of stellar magnetic fields. When measuring the magnetic field directly, we rely on the Zeeman effect on spectral lines, either from intensity spectra or spectropolarimetry \citep[see e.g.][for a review]{donati:2009}. These different methods can characterise magnetic fields on different spatial scales on cool stars, due to the signal cancellation in the spectropolarimetric observations caused by nearby surface regions with opposite magnetic polarities.

The variation of large-scale magnetic fields has been studied extensively using spectropolarimetric observations. Either by using longitudinal magnetic field measurements or complete magnetic field geometries from Zeeman Doppler imaging \citep[ZDI, see e.g.][]{kochukhov:2016}. Using these methods, both short-term variations, such as rotational modulation, and long-term evolution, such as polarity reversals of the magnetic field geometry caused by the magnetic cycle \citep[e.g.][]{boro-saikia:2018, bellotti:2025}, have been observed. 

In addition to the large-scale magnetic field, there are also plenty of unresolved magnetic features on the stellar surface. This small-scale field would also include starspots and faculae, as their size is insufficient to be resolved by ZDI, especially when accounting for a random distribution of magnetic field polarities that will cancel out any polarimetric signals. This means that whenever measuring magnetic fields using polarimetry, contributions from many of the magnetic field structures will not be recovered. In fact, when comparing the total and large-scale magnetic field strengths of Sun-like stars, \cite{kochukhov:2020a} found that less than $\sim10$\,\% of the magnetic field strength is recovered from ZDI, {showing that the small-scale magnetic field is the dominant component for these stars}.
It is possible to obtain the {total unsigned} magnetic field strength\footnote{The term `small-scale' field strength is commonly used in the literature to refer to the total unsigned magnetic field strength. While the small-scale magnetic field {generally} contributes to {over} $99\%$ of the magnetic field energy at the surface of {Sun-like} stars, using this term may lead to misunderstandings. Therefore, we have chosen to use the term `total unsigned magnetic field' throughout this paper.}
on stellar surfaces by {applying the Zeeman broadening technique to} unpolarised spectra \citep[e.g.][]{robinson:1980}. This is done with a combination of broadening and intensification of spectral lines caused by the magnetic field present on the stellar surface. While not sensitive to the magnetic field direction, this diagnostic provides information about 
the tangled and small-scale 
surface magnetic field structures, as the effect is not subject to the same signal cancellation as spectropolarimetric measurements. The temporal variation of the total unsigned magnetic field has, until recently, been studied in less detail compared to the large-scale field as multiple spectra are commonly averaged before performing the {Zeeman broadening} analysis \citep[e.g.][]{shulyak:2019,Reiners:2022}. Snapshot observations have hinted at statistically significant variations over timescales of weeks to years \citep[e.g.][]{hahlin:2023,pouilly:2024}. Time-series observations covering both rotational and longer timescales have shown indications of both rotational modulations \citep[e.g.][]{kochukhov:2017,donati:2023} and long-term evolution \citep[e.g.][]{bellotti:2023} of the {total unsigned} magnetic field. These clear detections of small-scale variability have, up until recent work on a set of M dwarfs by \cite{cristofari:2025}, been limited to very active M dwarfs. Furthermore, \cite{lavail:2019} found no strong signs of rotational modulation on T Tauri stars. For more moderately active Sun-like stars, \cite{kochukhov:2020a} found no significant variations from optical spectra between different observational epochs, and \cite{hahlin:2023} reported variations from snapshot observations over the time span of a few months using H-band observations. Furthermore, \cite{petit:2021} found no rotational modulation of $\varepsilon$ Eridani while \cite{lehmann:2015} found a statistically significant evolution of the small-scale magnetic field with an approximately 3-year cycle.

The surface-averaged small-scale magnetic field on the Sun, as calculated by disc integrating the line-of-sight magnetic field from SDO images \citep[see][for details]{haywood:2016}, shows that it constantly changes due to the rotation and evolution of active regions. Furthermore, the disc-integrated {total unsigned} magnetic field on the Sun appears to trace effects such as radial velocity variations, making the small-scale field a promising tool to improve our precision in radial velocity surveys. However, only the Sun has a surface sufficiently resolved for the magnetic field strength of individual active regions to be measured. To apply this approach in other stars, the small-scale magnetic field would have to be extracted from disc-integrated stellar spectra.

To better understand how magnetic fields influence stellar properties, it would be interesting to characterise the variation of the small-scale magnetic field in more detail. This could help in better understanding the role of the small-scale field on stars other than the Sun, as well as how it interacts with other phenomena used to study stellar activity. Another point of interest is to look at how different spatial scales of the stellar magnetic field are connected to each other. By quantifying how the large- and small-scale magnetic fields are correlated, we would be able to better understand how the underlying small-scale field structures influence the observed magnetic geometry measured from polarised observations. 

To investigate the variability of small-scale magnetic fields, we selected a collection of stars with already well-characterised large-scale magnetic fields from \cite{folsom:2016,folsom:2018}. We focus on the stars TYC 6349-0200-1 (HD 358623), HIP 76768 (HD 139751), HH Leo (HD 96064, HIP 54155), and Mel 25-5 (HIP 16908). The selection was based on two criteria. The first was that they showed clear and relatively simple rotational modulation of the longitudinal magnetic field. Secondly, the stars were also selected to cover the complete age range ($\sim10$ -- $600$\,Myr) studied in \cite{folsom:2016,folsom:2018}. This allowed us to test the possibility of recovering {total unsigned} magnetic field variations on a range of active stars. As the time series used for this study are generally limited to a few weeks to about a month, this study primarily examines variations over a few rotational phases. While longer variations of the {total unsigned} magnetic field are obviously of interest to monitor how the small-scale field evolves over a stellar activity cycle \citep[e.g.]{bellotti:2023,lienhard:2023}, the shorter time window should allow us to study the total magnetic field on the stellar surface without significant evolution of the structures present on the surface.

The paper is organised as follows: The spectroscopic observations are described in Sect.~\ref{sec:obs}, including additional data reduction steps. In Sect.~\ref{sec:small-scale}, the process for determining the {total unsigned} magnetic field on the stellar surface is outlined and the results for each star are presented. In Sect.~\ref{sec:discussion} we discuss the variability, or lack thereof, in the {total unsigned} magnetic field measurements. Then the field measurements are examined in the context of other activity indicators and in the context of the variability of other activity indicators. The results of this work are summarised in Sect.~\ref{sec:summary}.

\section{Observations}
\label{sec:obs}
\subsection{ESPaDOnS and NARVAL spectroscopy}
\begin{table*}
    \centering
    \caption{Observation log.}
    \label{tab:observations}
    \begin{tabular}{llccrr}
        \hline\hline
        Target &Instrument& Time window & N$^*$ & S/N$^{**}$ \\
        \hline
        TYC 6349-0200-1 & ESPaDOnS & 2013 Jun 15--30 & 16 & 100--120\\
        HIP 76768 & ESPaDOnS & 2013 May 18--30 & 24 & 40--130 \\
        \multirow{2}{*}{HH Leo} & NARVAL & 2015 Mar 6--May 26 & 14 & 100--300\\ 
         & NARVAL & 2017 Mar 28--Apr 21 & 14 & 220--280\\ 
        Mel 25-5 &  ESPaDOnS & 2015 Sep 18--Oct 1 & 14 & 140--200\\
        
        \hline
    \end{tabular}
    \tablefoot{$^*$: Number of observations obtained for each star and epoch. $^{**}$: Median S/N per pixel in the region used for magnetic field measurements, {see Table~\ref{tab:activity_data} for full set of observations}. 9660--9800\,\AA\,for HIP 76768, TYC 6349-0200-1, and Mel 25-5, and 5400--5520\,\AA\, for HH Leo.}
\end{table*}
The observations analysed in this study were obtained with ESPaDOnS\footnote{Data obtained from \url{https://www.cadc-ccda.hia-iha.nrc-cnrc.gc.ca/en/cfht/}} at the Canada France Hawaii Telescope and NARVAL at the Bernard Lyot Telescope. Both spectropolarimeters have a resolving power of $\sim$68000, cover the wavelength range between 3700 and 10000\,\AA, and are capable of observing any of the four Stokes parameters for spectropolarimetric analysis. The data were automatically reduced by the \textsc{Libre-ESpRIT} package \citep{donati:1997} and has been used previously (with the exception of the 2017 data of HH Leo) to measure activity indices (the S index, H$\alpha$ index, and Ca IRT index), longitudinal magnetic fields, and surface magnetic field distributions using ZDI of the stars \citep[see][]{folsom:2016,folsom:2018}. The information about the observations of each star can be found in table~\ref{tab:observations}. 

The ESPaDOnS observations were obtained during a time window of $\sim$2 weeks, which means that any variation in magnetic field strength or activity signals detected in these time series is unlikely to originate from any long-term evolution of the magnetic field on the stellar surface due to cycles or other forms of activity. The data from NARVAL of HH Leo was obtained over a longer time frame, almost three months in 2015 and about one month in 2017. For this reason, these observations could exhibit a larger evolution of the surface field within the time series. With the exception of HH Leo, the cadence of observations ranges from about two times per night to every other night.

\subsection{Telluric correction and continuum normalisation}
For the {total unsigned} magnetic field analysis of most stars, as described in Sect.~\ref{sec:small-scale}, we utilise a \ion{Ti}{I} multiplet with lines located between 9600--9800\,\AA. Unfortunately, this region is significantly affected by telluric absorption, making the lines challenging to use without some form of telluric removal. We use \textsc{Molecfit} \citep{smette:2015} to remove the telluric contamination. Following the procedure from \cite{kochukhov:2017}, we modify the fits headers for compatibility with \textsc{Molecfit} and perform the telluric removal. For this step, we only input the wavelength region between 9630--9900\,{\AA} of the observations.

We also performed an additional continuum normalisation of all spectra using a third-order polynomial. At this point, we also manually investigate the spectra for any issues in the telluric removal. We find that some spectra still have significant telluric residuals in some lines, which results in a rejection of a few spectra for some observation sequences (between 0 and 3 for each star). As the method used for the total magnetic field measurements used in Sect.~\ref{sec:small-scale} is dependent on high-quality spectra, we also rejected a handful of observations from the variability analysis with median S/N $<70$ in the wavelength region\footnote{Note that the S/N in these regions are generally lower than the values reported in \cite{folsom:2016,folsom:2018}} around the lines used for magnetic field investigations. This is motivated by the fact that at these S/N levels, the magnetic field results became increasingly unstable depending on the input parameters. In some cases, this could create outliers that bias the investigation of variability. With these rejection steps, all stars and epochs still have a sufficient number of observations to provide a complete phase coverage according to the reported periods. In addition, including the rejected spectra in the analysis appears to have little influence on the overall periodic modulation reported on each star.

\section{Total unsigned magnetic field inference}

\begin{table*}
    \centering
    \caption{Stellar parameters}
    \label{tab:stellar-parameters}
    \begin{tabular}{llcccccrr}
        \hline \hline
        Target & \multicolumn{1}{c}{$T_{\rm eff}$ (K)} & $\log g$ & $v_{\rm mic}$ (km\,s$^{-1}$) & {$R_{\star}$ ($R_\odot$)} & Period (d) & Age (Myr)& $\langle B_{\rm ZDI}\rangle$ {(G)} & Source \\
        \hline
        TYC 6349-0200-1 & $4359\pm131$ & $4.19\pm0.31$ & $1.4\pm0.3$ & $0.96\pm0.07$ & $3.41\pm0.05$& $24\pm3$ & 59.8 & 1,2 \\ 
        HIP 76768 & $4506\pm153$ & $4.53\pm0.25$ & $0.6\pm0.3$  & $0.85\pm0.11$ & $3.70\pm0.02$ & $120\pm10$ & 112.8 & 1,2\\
        HH Leo & $5402\pm73$ & $4.62\pm0.11$ & $1.31\pm0.31$  & $0.84\pm0.03$ & $5.915\pm0.017$ & $257\pm46$ &13.0 & 3 \\
        Mel 25-5 & $4916\pm97$ & $4.35\pm0.22$ & $1.01\pm0.19$  & $0.91\pm0.04$ & $10.57\pm0.10$ & $625\pm50$ & 28.9 & 3,4 \\
        \hline
    \end{tabular}
    \tablefoot{1: \cite{folsom:2016}, 2: \cite{messina:2010}, 3: \cite{folsom:2018}, 4: \cite{delorme:2011}}
\end{table*}

\label{sec:small-scale}

We aim to measure how the properties of the surface magnetic field from disc-integrated intensity spectra vary over time. This property is obtained from the distortion of spectral lines caused by the Zeeman effect. In contrast to polarised observations, the signal in the intensity spectra is insensitive to the orientation of the magnetic field \citep[see e.g.][]{kochukhov:2021}. While this means that we do not gain information about the geometry of the local field, it also means that we do not suffer from any signal cancellation of nearby surface elements. This effectively means we can recover the complete magnetic field strength on the stellar surface. This includes field structures that are not resolvable with ZDI, such as starspots. 

We use the magnetic inference code from \cite{hahlin:2023} to determine the total magnetic field. It uses Markov chain Monte Carlo (MCMC) sampling from the {Solar Bayesian Analysis Toolkit}\footnote{\url{https://github.com/Sergey-Anfinogentov/SoBAT}} \citep[SoBAT;][]{anfinogentov:2021} to find both magnetic and non-magnetic parameters simultaneously. The non-magnetic free parameters used are the chemical abundance (given as $\log (N/N_{\rm total})$), non-magnetic broadening (projected rotational velocity $v\sin{i}$ or macroturbulence $v_{\rm mac}$), and radial velocity. For the magnetic field, we use a multi-component model to describe the surface distribution of the magnetic field, this means that the surface magnetic field consists of a fixed set of magnetic regions, evenly distributed over the stellar disc. 
The total unsigned magnetic field ($B_I$) at the stellar surface is then determined from,
\begin{equation}
    B_{I}=\sum_{i}f_{i}B_{i}.
\end{equation}
With $f_i$ representing the filling factor of each magnetic field strength used in the model.
We set the strengths of these regions to be steps of 2~kG. Another aspect of this code is that the error can also be used as a free parameter, this is justified by the fact that in high S/N studies using high resolution spectroscopy the residuals are often dominated by systematic effects rather than noise. By including the error as a free parameter, we can give the inference more freedom to find a suitable fit. The consequence of implementing this is primarily more conservative error estimates while the median parameters tend to be unaffected.

As the code has previously been used on individual observations, we have made some modifications in order to treat a time series with more consistency. This involves optionally fixing non-magnetic parameters such as abundance and non-magnetic broadening to not let them vary across observations in the same time-series. As these parameters should be stable, any variation in non-magnetic parameters could hide the magnetic variation. In order to constrain the non-magnetic parameters, we use the time averaged spectra, find the best fit parameters {(both magnetic and non-magnetic)}, and then fix the abundance and non-magnetic broadening parameters for each individual spectrum. In order to limit the number of filling factors used, we use the Bayesian information criterion \citep[BIC;][]{sharma:2017}, to weight the improvement of fit with the increased complexity of adding additional filling factors. The BIC has the following form,
\begin{equation}
\label{eq:bic}
    \text{BIC} = -2\ln{p(Y|\hat{\theta})}+d\ln{n},
\end{equation}
where $p(Y|\hat{\theta})$ is the best fit likelihood, $d$ is the number of parameters and $n$ is the number of data points. We select the model producing the lowest BIC value with the motivation that any increased complexity beyond the lowest BIC will not have a significant improvement to the fit.

As most of our stars have $T_{\rm eff}$ in the 4000--5000\,K range, we elect to primarily use the magnetically sensitive multiplet of \ion{Ti}{I} lines in our study. These lines have several advantages as outlined in \cite{kochukhov:2017}. First, the lines being from the same multiplet means that the relative oscillator strengths are well-known. 
Second, there is a large range of sensitivities to magnetic fields, the multiplet has a magnetic null line (a spectral line not affected by the magnetic field) making the lines ideal to constrain non-magnetic parameters simultaneously with the magnetic field.

The synthetic spectral grid is generated with line lists from VALD \citep{ryabchikova:2015}, MARCS model atmospheres \citep{gustafsson:2008} and the polarised radiative transfer code \textsc{Synmast} \citep{kochukhov:2007}. We use stellar parameters $T_{\rm eff}$, $\log g$, $v_{\rm mic}$ for each star given by \cite{folsom:2016} or \cite{folsom:2018} as shown in Table~\ref{tab:stellar-parameters}. 
The grid is calculated before the MCMC sampling begins, where we fix $T_{\rm eff}$ and $\log g$ using bi-linear interpolation within the MARCS grid. 
We generate the spectral grid using abundances in a range of $\pm 0.15$ from an initial test fit with a step size of 0.05 and magnetic field strengths in steps of 2\,kG. 
Each generated spectra is then disc-integrated for different strengths of non-magnetic broadening. Here, the non-magnetic broadening parameter has a step size of 0.5\,km\,s$^{-1}$ in the grid.
The macroturbulent velocity is determined by the empirical relationship from \cite{doyle:2014} (unless it is used as a free parameter) and $v\sin{i}$-values are taken from \cite{folsom:2016,folsom:2018}. While both HIP 76768 and TYC 6349-0200-1 are outside of the validity range of the empirical relationship for macroturbulence, their high $v\sin{i}$-values mean that $v\sin{i}$ is the dominant process for the line shape. In either case, changing macroturbulence only influences the optimal $v\sin{i}$-value without any significant effect on the average magnetic field. The resulting median parameter values from the obtained posterior distributions for each star can be seen in Table~\ref{tab:average}. The fit and corner plot for one case (TCY 6349-0200-1) are shown in Appendix~\ref{app:repfit}  

\begin{table*}[]
    \centering
    \caption{Time-averaged inference parameters}
    \begin{tabular}{lccc}
        \hline\hline
        Star & $\varepsilon_{\rm Ti}$ & $v\sin{i}$ (km\,s$^{-1}$) & $\langle B_I\rangle$ (kG) \\
        \hline
        HIP 76768 & $-7.21\pm0.02$ & $10.29\pm0.21$ & $2.73\pm0.11$ \\
        TYC 6349-0200-1 & $-7.15\pm0.01$ & $15.41\pm0.13$ & $1.212\pm0.075$ \\
        \hline
         & $\varepsilon_{\rm Ti}$ & $v_{\rm mac}$ (km\,s$^{-1}$) & $\langle B_I\rangle$ (KG) \\ 
        \hline
        Mel 25-5 & $-7.14\pm0.10$ & $2.61\pm0.21$ & $0.732\pm0.065$ \\
        \hline
        & $\varepsilon_{\rm Fe}$ & $v\sin{i}$ (km\,s$^{-1}$) & $\langle B_{I}\rangle$ (kG)\\
        \hline
        HH Leo (2015) & $-4.86\pm0.01$ & $6.95\pm0.10$ & $0.424\pm0.052$ \\
        HH Leo (2017) & $-4.86\pm0.01$ & $6.98\pm0.11$ & $0.498\pm0.052$\\
        \hline
    \end{tabular}
    \tablefoot{Slightly different parameter choices were made for some stars, check individual sections for details.}
    \label{tab:average}
\end{table*}

\subsection{HIP 76768}
\label{sec:hip}
\subsubsection{Time-averaged spectra}
\label{sec:hip_avg}

We produce a time-averaged spectrum from the ESPaDOnS observations by combining the time series observations {described in Sect~\ref{sec:obs}}. This was done by cross-correlating each spectra with a reference spectra to identify any radial velocity offsets between different observations. Then, we co-added the spectra, using a weighted mean, into the same wavelength grid by shifting each spectra with its radial velocity. As \textsc{Libre-ESpRIT} shifts the spectra into a barycentric rest frame, the cross-correlation has a marginal effect on the output spectrum. It does, however, ensure that we correct for other radial velocity variations in the observations.

Using the stellar parameters from Table~\ref{tab:stellar-parameters}, we generate a synthetic grid and use MCMC sampling to find the magnetic filling factors, abundance, and rotational velocity of HIP 76768. Comparing different numbers of filling factors, in the case of HIP 76768 we explored all models up to 12~kG. We find that the BIC {(eq.~\ref{eq:bic})} favours a rather complex model {(see Table~\ref{tab:BIC} for values)} consisting of six components with field strengths of 0, 2, 4, 6, 8, and 10\,kG. {The magnetic field obtained for HIP 76768 from the time-averaged spectrum, labelled as $\langle B_I\rangle$, was found to be} 2.7~kG.

\subsubsection{Verification of stellar parameters}

In order to verify that our choice of stellar parameters from \cite{folsom:2016,folsom:2018} does not significantly influence our results, we re-run the MCMC inference for the mean spectrum of HIP 76768 while also including effective temperature and surface gravity as additional free parameters. 
Our initial guess is the median parameters of the original analysis {in Sect.~\ref{sec:hip_avg}}. We use uniform priors for both $T_{\rm eff}$ and $\log g$. 
The procedure is the same, except that we expand the $T_{\rm eff}$ and $\log g$ grid to the neighbouring points in the MARCS grid. The median values, as well as the difference between the original analysis, can be seen in Table~\ref{tab:param_diff}. 

\begin{table}
    \centering
    \caption{Comparison between obtained stellar parameters for HIP 76768}
    \label{tab:param_diff}
    \begin{tabular}{llrrr}
        \hline\hline
        \multicolumn{2}{l}{Parameter} & Value & $\Delta$ & $\Delta/\sigma^{**}$ \\
        \hline
        $T_{\rm eff}$$^{*}$ & (K) & 4506 & 6 & 0.04\\
        $\log g$$^{*}$ & & 4.53 & 0.13 & 0.52\\
        $\varepsilon_{\rm Ti}$ & & -7.27 & -0.06 & --\\
        $v\sin i$ & (km\,s$^{-1}$) & 10.22 & -0.07 & -0.12\\
        $\langle B_{I}\rangle$ & (kG) & 2.85 & 0.12 & 1.05\\
        \hline
    \end{tabular}
    \tablefoot{Reference values taken from Table~\ref{tab:stellar-parameters} and \ref{tab:average}. $^*$ Parameter was held fixed in the original analysis. $^{**}$ $\sigma$ is taken from table 2 of \cite{folsom:2016} except for the $\langle B_{I}\rangle$ case where it is taken from this work.}
\end{table}

From the result, we can see that there is a close agreement between the obtained and adopted stellar parameters. Both $T_{\rm eff}$, $\log g$, and $v\sin i$ are all within the uncertainties reported for HIP 76768 by \cite{folsom:2016}. We note that our own uncertainties for $T_{\rm eff}$, $\log g $, and $\varepsilon_{\rm Ti}$ are quite low. This is a common occurrence for the abundance parameter in other magnetic field studies \citep[e.g.][]{hahlin:2023} and likely comes from the fact that all lines investigated here are from the same multiplet and therefore share a similar response to changing stellar parameters. The good agreement between standard spectroscopic analysis methods and methods including a proper magnetic field description has also been found by \cite{cristofari:2023}, showing that magnetically sensitive lines can be used for stellar characterisation provided the magnetic field effects are accounted for.
The magnetic field value changes by slightly above 0.1\,kG, this is a variation of about $1\sigma$ compared to the case when $T_{\rm eff}$ and $\log{g}$ were held fixed. This reflects the results of other investigations of magnetic sensitivity to stellar parameter choice \citep[e.g.][]{kochukhov:2020a, hahlin:2022, hahlin:2023} that find similar variations.

We find that when we fit models with fewer filling factors, obtained parameters can be quite different. For example, when using a model with magnetic field strengths of 0, 2, and 4~kG, the obtained temperature is 4660~K. {However, the $\chi^2$ is a factor of 3.6 larger compared to the 6-component model. This shows that the model is significantly worse at fitting the data}. In addition, if the magnetic field is not included in the analysis of magnetically sensitive lines, there is a significant risk of introducing systematic errors into fundamental stellar parameters.

While we are able to obtain non-magnetic parameters self-consistently with our MCMC inference, we elect not to use it for the entire sample. The reason is the increased computational cost, {in part due to} adding two extra dimensions to the spectral grid. {In addition,} we found that the convergence time of the MCMC sampling for adding $T_{\rm eff}$ and $\log g$ significantly increased compared to when they were held fixed. Given that our results for HIP 76768 are consistent with literature values, we therefore elected to use the established values for these parameters for our analysis.

\subsubsection{Individual observations}

With the non-magnetic median parameters from Table~\ref{tab:average}, we then determine the magnetic parameters of each individual observation with the non-magnetic parameters fixed {see Table~\ref{tab:activity_data}}. For consistency, we use the same number of filling factors as in the mean spectrum case. The field strengths obtained for each observation lies in the range 2.6 -- 3.3 kG, this variation is significant compared to the $\sim0.1$~kG uncertainties which indicates that there are substantial variations in the disc averaged magnetic field strength over a time period of a few weeks. 

\subsection{TYC 6349-0200-1}
\label{sec:tyc}
The analysis of TYC 6349-0200-1 is carried out with the same procedure as described for HIP 76768 in Sect.~\ref{sec:hip}. For this star, we find a field strength from the time averaged spectra of around 1.2\,kG with individual measurements in the range between 1.1 -- 1.3\,kG, this variation does not appear to be significant. Besides the weaker field, we also find that a simpler model is more favoured for the filling factors, specifically a model with 4 components with strengths of 0, 2, 4, and 6\,kG. 

The intermediate filling factor is consistent with 0 (see Fig.~\ref{fig:corner}). While this could indicate a surface with highly variable magnetic field strengths, the correlation between $f_4$ and the other filling factors also indicates that there is cross-talk between the filling factors. In fact, the rotational broadening of TYC 6349-0200-1 is significantly larger ($\sim15$\,km\,s$^{-1}$) compared the broadening of spectral lines caused by the Zeeman effect ($\sim 5$\,km\,s$^{-1}$ for a 2\,kg field). This means that we are more sensitive to intensification effects related to the strength of the lines, rather than the detailed line shape, making individual filling factors more difficult to infer. While not favoured by the BIC (see Table~\ref{tab:BIC}), we find a similar behaviour from the three-component model where the $f_2$-component is consistent with zero and most of the contribution comes from the $f_4$-component. This further highlights the difficulty to disentangle the filling factors at larger rotation rates as different choices of filling factor can produce very different results without substantially changing the magnetic field.

\subsection{Mel 25-5}
For Mel 25-5, we elect to use $v_{\rm mac}$ as the free parameter for non-magnetic broadening. The reason for this is that Mel 25-5 has the lowest $v\sin{i}$ in the sample at 3.28~km\,s$^{-1}$ \citep{folsom:2018}, this means that the rotational broadening is likely comparable to the broadening caused by macroturbulence. As $v\sin{i}$ is more closely tied to the geometry and period of the star, we elect to keep it fixed {at 3.28\,km\,s$^{-1}$ from \cite{folsom:2018}}. Instead, we let $v_{\rm mac}$ be the free parameter. Other than that, the analysis is performed in the same way as described in Sect.~\ref{sec:hip}. For this star, we find a magnetic field strength of $\sim0.7$\,kG using a multi-component model with 0, 2, 4, and 6\,kG. The individual measurements have a variation between 0.6 -- 0.8\,kG, which is a bit above two $\sigma$ between peak to peak.

\subsection{HH Leo}
\label{sec:hhleo}
HH Leo is the hottest star in the sample by about 500\,K compared to the other stars. {This is a problem when using the \ion{Ti}{I} lines for magnetic inference. While these lines have been used extensively on stars with spectral types M and K, they are not commonly used for hotter stars. In the case of HH Leo the lines are much weaker compared to the other stars in this study}. For this reason, we instead choose to use the magnetically sensitive \ion{Fe}{I} lines identified by \cite{kochukhov:2020a}. As these lines are located at around 5500\,\AA\, within the optical, there is no significant telluric contamination, which means that we do not need to process the spectra using \textsc{Molecfit}. Similarly to the \ion{Ti}{I} lines, these lines originate from the same multiplet, have a range of magnetic sensitivities, and include a line that is insensitive to the magnetic field. 

Besides the choice of magnetically sensitive lines, the star was also observed twice. One of the datasets is from 2015 and was used in \cite{folsom:2018} to measure the large-scale magnetic field. Another, unpublished data set from 2017 was also available. To explore variations on larger timescales, we analyse these two sets separately by making a time-average spectrum of each epoch individually and then perform two independent analyses with these two averaged spectra. This yields field strengths from the time-averaged spectra of $0.43$ and $0.50$\,kG for the 2015 and 2017 epochs, respectively. In both cases the favoured model contained three components with strengths of 0, 2, and 4\,kG. Given the uncertainty of $\sim0.05$\,kG, this epoch variation is slightly significant. For the individual measurements, both epochs have peak-to-peak variations of $\sim0.2$\,kG around the corresponding average result. While this indicates potential variation over the two years, the scatter of individual measurements is sufficiently large to create a substantial overlap. We also note that, even though the two epochs are fitted independently, the non-magnetic parameters obtained from the two epochs are in agreement with each other.

One notable aspect with the HH Leo results is that there is little variation in the BIC-values (see Table~\ref{tab:BIC}). When testing other model comparison parameters \citep[AIC and WAIC, see][]{sharma:2017}, we find that there is not a consistent agreement of which model is preferred between them. This indicates that neither the two- or three-component is strongly preferred. We elect to persist with the BIC criterion posed for the other stars for consistency, and note that the obtained $\langle B\rangle$-values for the two-component model is within 0.01\,kG of the three-component model. The model choice would therefore have a negligible impact on the result. One benefit with the three-component model is that the degeneracy between the magnetic and non-magnetic parameters are reduced, without changing the median values.

\section{Discussion}
\label{sec:discussion}
\subsection{Total unsigned magnetic field variations}
As some of the stars exhibited statistically significant variations in the total unsigned magnetic field measurements, it is worthwhile to explore this in more detail. To check for rotational modulation, we fit a sinusoidal function to our observed data where the frequency is fixed by the rotational period of the star from the references in Table~\ref{tab:stellar-parameters}. To see if the period is recoverable on its own, we also calculate a Lomb-Scargle periodogram \citep{zechmeister:2009} on the same time series.

\subsubsection{HIP 76768}

\begin{figure*}
    \centering
    \includegraphics[width=0.45\linewidth]{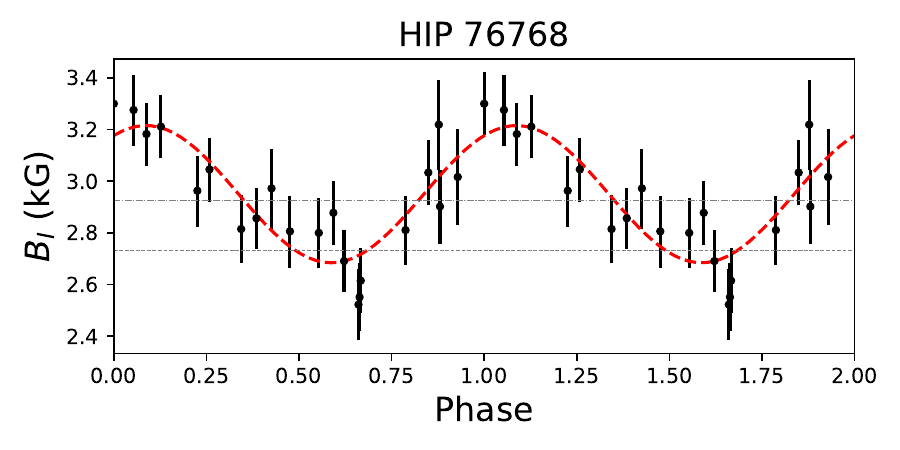}
    \includegraphics[width=0.45\linewidth]{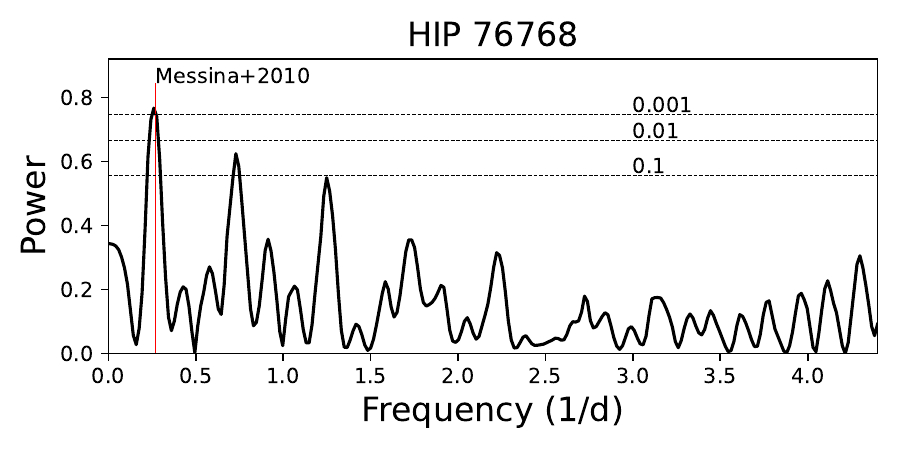}
    \caption{Magnetic field periodicity of HIP 76768. Left: Total unsigned magnetic field variation of HIP 76768 phase-folded using the rotational period from \cite{messina:2010}. The dashed red line represents the best-fit sinusoidal function to the data with the fixed rotational period from Table~\ref{tab:stellar-parameters}. The dashed grey line represents the average of the individual measurements, while the dotted line represents the {total unsigned} magnetic field obtained from the time-averaged spectrum. Right: Lomb-Scargle periodogram obtained from the magnetic field measurements of HIP 76768. The dashed lines indicate false alarm probabilities. Also shown is the expected period from Table~\ref{tab:stellar-parameters}. Similar plots for all stars can be seen in Fig.~\ref{fig:magnetic_variability}.}
    \label{fig:hip-variation}
\end{figure*}

The phase-folded result and the sinusoidal fit to the HIP 76768 magnetic field strength can be seen in Fig.~\ref{fig:hip-variation}. From this, we can clearly see that the measured magnetic field strength follows the rotational phase of the star. When comparing the variation with the field from the average spectra, we see that the average spectra produces a slightly weaker field strength compared to most of the individual measurements, although still within the observed variation. When looking at the fit quality, we find that the reduced-$\chi^2$ is around 0.7, this could indicate that the sinusoidal function is over-fitting the trend, or that we have slightly overestimated the uncertainties by making the observational errors a free parameter in our inference.

This periodicity is further supported by the Lomb-Scargle periodogram shown in Fig.~\ref{fig:hip-variation}. Here we see a peak in the periodogram at a frequency that matches the period from \cite{messina:2010} adopted by \cite{folsom:2016}. The other strong peaks appear to be multiples of the frequency corresponding to the strongest peak (i.e. harmonics). This shows that, while the sinusoidal fit might be slightly over-fitted, there is a clear preference for the magnetic field to vary with a frequency corresponding to the rotational period of the star. As these observations were obtained within about 12 days, or three rotational cycles, it is unlikely that any other significant magnetic field evolution would have happened during the observations. This not only points to the fact that the total magnetic field can be significantly different in different regions on the stellar surface but also that the rotational period can be recovered self-consistently from the magnetic field measurements.

\subsubsection{TYC 6349-0200-1}
The phase-folded results of TYC 6349-0200-1 can be seen in Fig.~\ref{fig:magnetic_variability}. As mentioned in Sect.~\ref{sec:tyc}, the variation of the individual measurements appeared not to be significant. This results in a phase-folded variation that does not show any strong rotational modulation. Upon closer investigation, it appears that the sinusoidal variation that we do see is primarily driven by the outlier point close to the 0.5 phase in Fig.~\ref{fig:magnetic_variability}. When that point is manually removed, the other data shows no sinusoidal signal as a function of phase. This is supported by the periodogram in Fig.~\ref{fig:magnetic_variability}, that shows no strong peak near the rotational period reported by \cite{folsom:2016}. 
Its lack of rotational modulation is similar to the results from \cite{lavail:2019} that found no rotational modulation on T Tauri stars. As TYC 6349-0200-1 is the youngest star in our sample, while showing the weakest magnetic field variation, our results suggest that young stars have a much more even distribution of magnetic fields compared to their more evolved counterparts. This is also supported by \cite{yamashita:2025}, which finds lower than expected longitudinal magnetic field variations in young stars compared to the main-sequence trend.

\subsubsection{MEL 25-5}
The phase-folded data of Mel 25-5, along with its sinusoidal fit, can be seen in Fig.~\ref{fig:magnetic_variability}. While the observational sampling is biased to phases between 0--0.5, it does appear to be a rotationally modulated variation in the measured magnetic field strengths. However, this signal is much weaker than in HIP 76768 and while there is a peak corresponding to the rotational period in the periodogram in Fig.~\ref{fig:magnetic_variability}, it has a large false alarm probability (>0.1). If the tentative rotational modulation is real, a big limitation for this star is likely the relatively large uncertainties of the magnetic field strengths, although our uncertainties might be slightly overestimated due to setting the error as a free parameter, as well as a phase-biased sampling.  
This indicates that for accurate constraints on the rotational modulation of magnetic fields on Mel 25-5, we would either need observations with higher S/N or a method able to constrain the magnetic field with higher precision. Alternatively, better sampling of the rotational period might be able to mitigate the noise limitations.

\subsubsection{HH Leo}
\begin{figure}
    \centering
    \includegraphics[width=\linewidth]{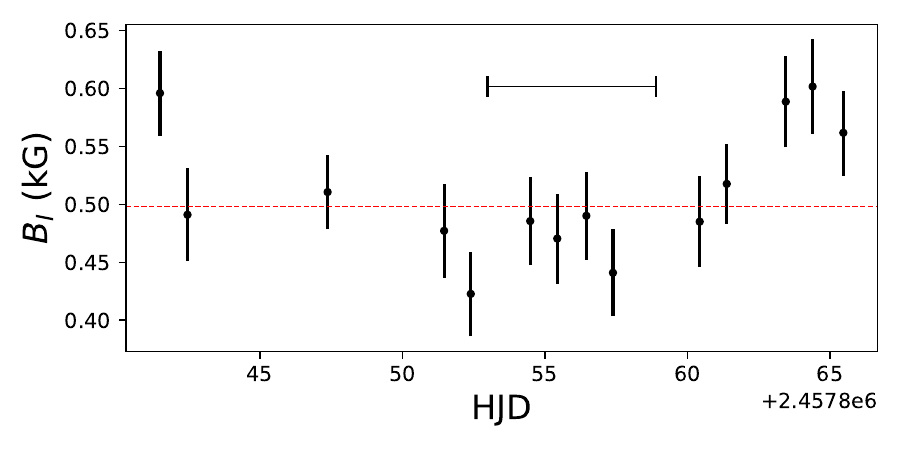}
    \caption{Total unsigned magnetic field measurements of HH Leo during 2017. The dashed red line marks the field strength obtained from the time-averaged spectra. The horizontal bar represents the rotational period.}
    \label{fig:HHLeo2017_epoch}
\end{figure}

We show the phase-folded data of the two observation epochs separately in Fig.~\ref{fig:magnetic_variability}. Although variations of a few $\sigma$ are detected, neither epoch shows any variation that would hint at a rotationally modulated magnetic field. One aspect that could obscure the rotational modulation of these targets is the longer time period over which the observations were obtained. Over a month or longer, cool stars could undergo sufficient changes to magnetically active regions of the surface to change the phase of any rotational modulation. Specifically, \cite{folsom:2018} reported a worse reduced $\chi^2$ for the 2015 ZDI map that was attributed to some evolution of the large-scale field even when accounting for differential rotation. In addition, the total magnetic field variation observed in the 2017 dataset appears to follow some trend as the strongest values are concentrated during the beginning and end of the observational epoch, as shown in Fig.~\ref{fig:HHLeo2017_epoch}. This could represent the evolution of surface structures, rather than rotational modulation, on the timescales of $\sim1$ month. However, no similar trends could be seen in the 2015 data.

\begin{figure}
    \centering
    \includegraphics[width=\linewidth]{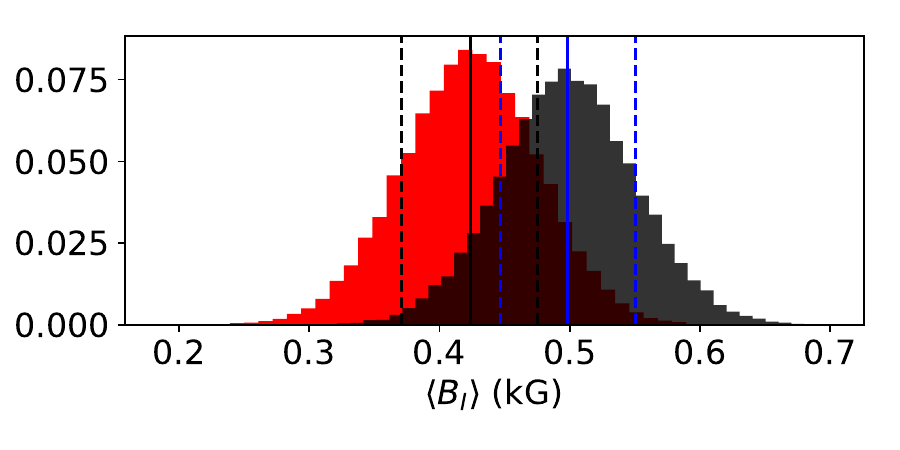}
    \caption{Posterior distribution of the  
    total unsigned magnetic field strengths obtained from the time-averaged spectra of HH Leo from the two epochs. The distribution from the 2015 data is shown in red, and the shaded region represents the 2017 data. The black and blue lines represent median and 68\% credence regions of the 2015 and 2017 datasets, respectively.}
    \label{fig:hhleo-comparison}
\end{figure}

The mean spectra of the two epochs appear to have a slight difference in the measured magnetic field strength at the 1$\sigma$-level, as shown in Fig.~\ref{fig:hhleo-comparison}. This variation is well within the scatter of each epoch but could hint at an evolution of the average activity level of HH Leo. Such trends have been detected on similar stars, such as $\xi$ Boo A where a tentative total unsigned magnetic field variation has been reported over a few years \citep[see][]{kochukhov:2020a,hahlin:2025}. 

\subsection{Comparison with other activity measurements} \label{sec:activity}
Many studies on the variability of stellar magnetic fields published over the past decades focused on indirect activity indices. While works such as \cite{Reiners:2022} have demonstrated a relationship between activity indicators and the total unsigned magnetic field for a large number of stars, the variation for an activity indicator at any given field strength can vary by an order of magnitude or more. Furthermore, their time-averaged spectra do not indicate if variations in an activity indicator correspond to a variation in the underlying magnetic field strength.

For the longitudinal magnetic field ($B_\ell$) and results from ZDI, similar results are reported by \cite{marsden:2014}, \cite{vidotto:2014}, \cite{brown:2022} and \cite{bellotti:2025a}, who find correlations between activity indicators, age, and magnetic field strengths with similar scatter as the total magnetic field measurements. \cite{see:2016} have also demonstrated that magnetic cycles determined from magnetic geometry can differ from activity cycles measured with other indicators.
From monitoring of individual stars, studies \citep[e.g.][]{amazogomez:2023,rescigno:2024} have found a lack of correlation between the longitudinal field and activity indicators. However, \cite{rescigno:2024} found that the RMS of $B_\ell$ over different time windows correlate well with the solar RV variations, indicating that large variations in $B_\ell$ on rotational timescales are connected to enhanced stellar activity. While activity and magnetic fields are clearly connected, whether these proxies show coherent variability with the magnetic flux over different timescales remains unclear. Evaluating this connection could help unveil the underlying mechanisms that drive stellar activity.

\begin{figure*}
    \centering
    \includegraphics[width=\linewidth]{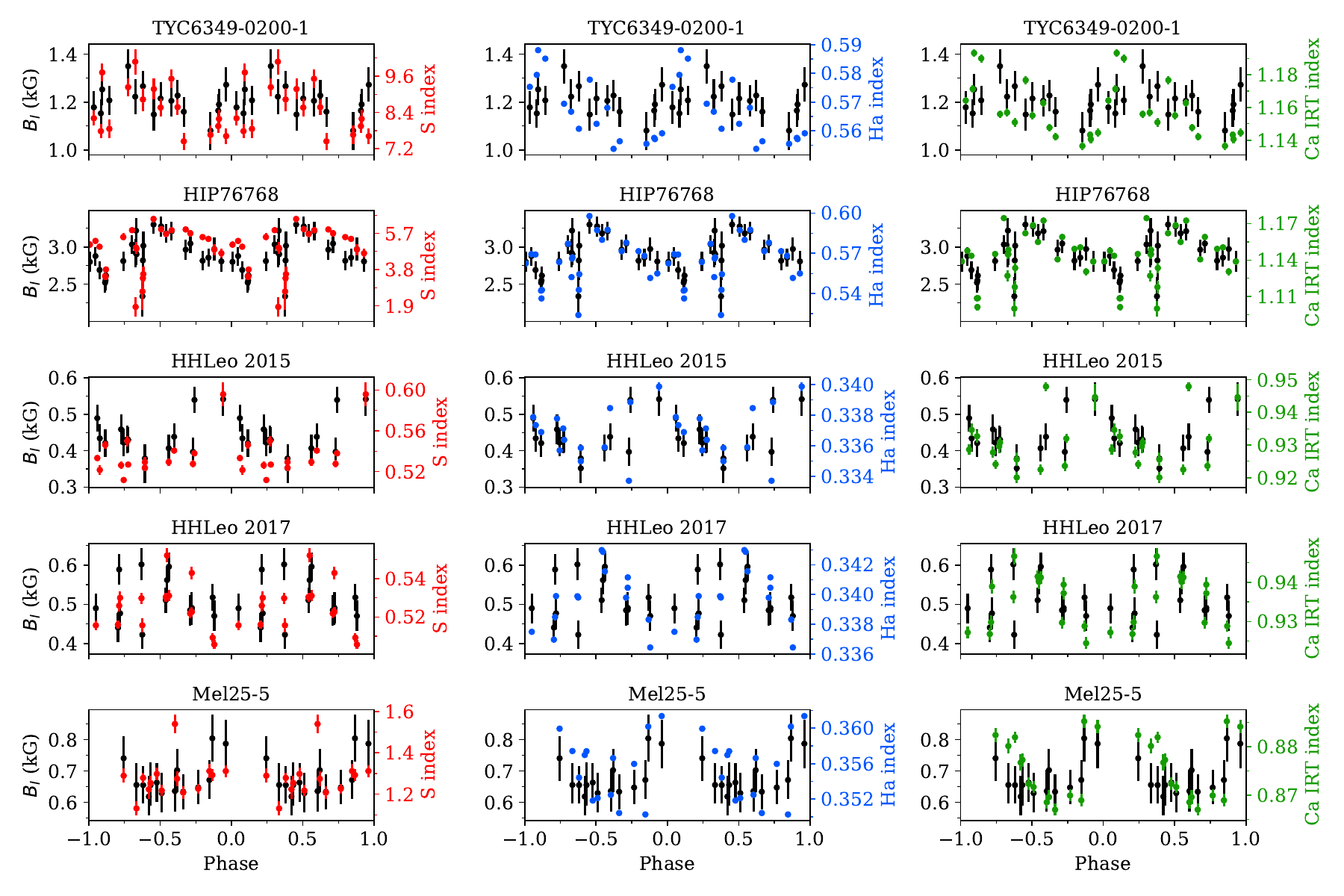}
    \caption{Comparison between the variation of total magnetic fields (left y-axis) and other activity indicators (right y-axis). From top to bottom rows display the results for the four targets studied with increasing age. From left to right columns show the results for the S index, H$\alpha$ index, and Ca IRT index.}
    \label{fig:activity-comparison}
\end{figure*}

Figure~\ref{fig:activity-comparison} illustrates the {temporal evolution of} the disc-integrated {total unsigned} magnetic fields derived in this study, {alongside} simultaneous activity indices reported by \citet{folsom:2016,folsom:2018}, namely the S index from \ion{Ca}{ii} H$\&$K, the H$\alpha$ index, and the Ca IRT index from the \ion{Ca}{ii} infrared triplet.
We observe a coherent variability, indicating that the temporal evolution of these activity indicators is closely linked to changes in the small-scale photospheric magnetic field. {As highlighted in Fig.~\ref{fig_extra:corr-small-large}, the total unsigned field is positively correlated with the activity traces for all stars but TYC 6349-0200-1. For this target, the Spearman correlation coefficient indicates a positive association only between $B_I$ and the S index.} 

\begin{figure}
    \centering
    \includegraphics[width=\linewidth]{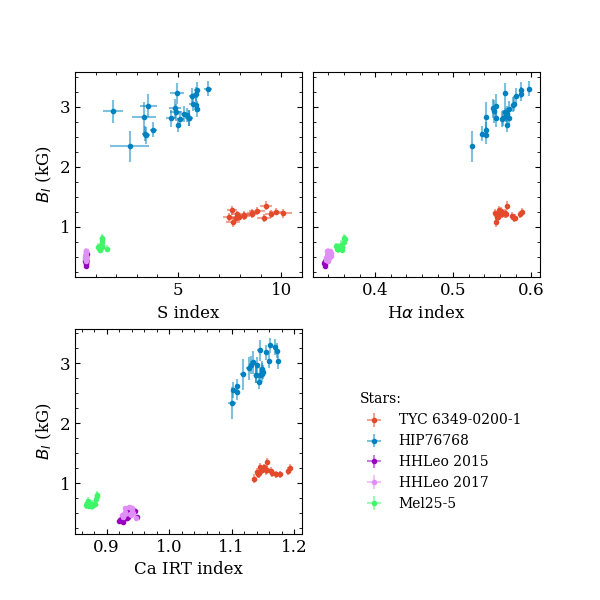}
    \caption{Correlations between the total unsigned magnetic field and other activity indicators. Top-left: S index. Top-right: H$\alpha$ index. Bottom-left: Ca IRT index. Symbol colours identify the four targets analysed in this study (see legend).}
    \label{fig:activity-correlation}
\end{figure}

Figure~\ref{fig:activity-correlation} presents the scatter diagrams of the relationship between magnetic field strength and activity indices, with symbol colours representing the measurements for the four targets analysed in this study. What stands out in this figure is that there is no linear relationship between $B_I$ and the activity indices when considering the full sample. This suggests that, while activity indices can effectively track temporal variations in the magnetic field for a given star, they cannot reliably determine its absolute magnetic field strength. A possible interpretation is that the way magnetic flux is distributed among plages, networks, and spots may significantly impact the S index, as suggested by the solar analysis of \citet{Cretignier:2024}.

These findings have important implications for studies that compare stars based on their activity indices. For example, H$\alpha$ and Ca IRT indices would indicate that TYC 6349-0200-1 and HIP 76768 have similar activity levels, whereas HIP 76768 has a significantly stronger magnetic field in reality. Similarly, the S index fails to recognise that HIP 76768 has a stronger magnetic field, erroneously suggesting that TYC 6349-0200-1 should be more magnetic. A similar discrepancy arises between HH Leo and Mel 25-5, where the Ca IRT index incorrectly identifies which star has a stronger magnetic field. It is worthwhile to note that we would see a similar behaviour if using the large-scale magnetic field strength from \cite{folsom:2016,folsom:2018} as the average $B_\mathrm{ZDI}$ of our targets (see Table~\ref{tab:stellar-parameters}) corresponds to between 5 and 3\,\% of our obtained $B_{I}$. 

Our results clearly indicate that activity indices are powerful tools for tracing magnetic variability over time and are, therefore, well suited to detect magnetic cycles. On the other hand, we also showed that relying solely on activity proxies to compare different stars can introduce significant biases, as they do not provide a comparable measure of the absolute stellar magnetic field strength. This limitation may contribute to the apparent spread observed in empirical relationships such as the magnetic activity–rotation relation \citep{Newton:2017,Astudillo:2017,Gomes:2021,Fritzewski:2021,Anthony:2022,Boudreaux:2022}, where part of the scatter could be spurious, arising not from intrinsic stellar properties (e.g., stellar mass) but from how magnetic flux manifests differently on stellar surfaces and translates into chromospheric emission. {As shown in Fig.~\ref{fig_extra:activity-correlationFlux}, similar conclusions can be drawn when considering how the unsigned magnetic flux, defined as $\Phi_I = 4\pi R_\star^2 B_I$, scales with the activity indices, given that the stellar radius does not vary significantly in the sample of stars examined (see Table~\ref{tab:stellar-parameters}).}

\subsection{Magnetic variations on different spatial scales}
\begin{figure*}
    \sidecaption
    \includegraphics[width=12cm]{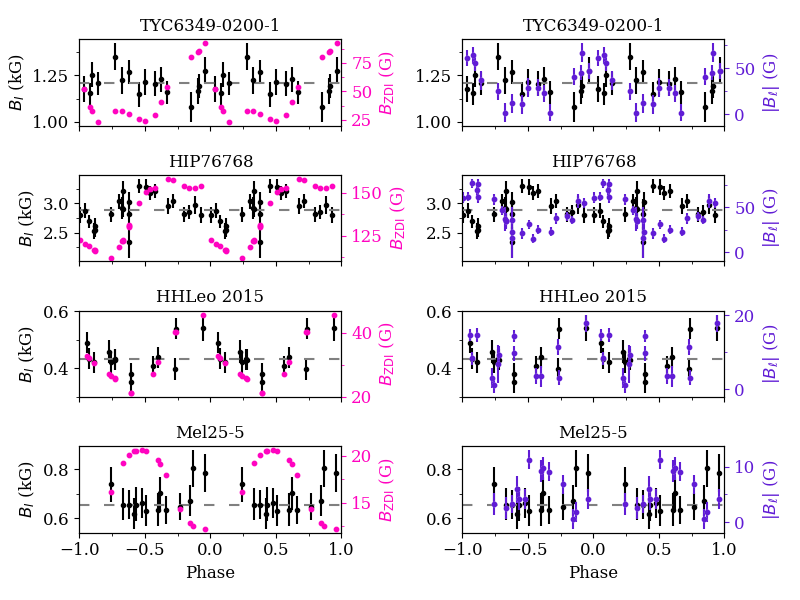}
    \caption{Similar to Fig.~\ref{fig:activity-comparison}, but for the comparison between the rotational variation of total unsigned magnetic fields and other large-scale polarimetric field measurements from \citet{folsom:2016,folsom:2018}. Left-column: Magnetic field strength obtained with Zeeman-Doppler Imaging technique. Right-column: disc-integrated longitudinal magnetic field modulus.}
    \label{fig:largescale}
\end{figure*}

Figure~\ref{fig:largescale} presents a comparison between the variability of the {total unsigned} magnetic field and two diagnostics of the large-scale field reported by \cite{folsom:2016,folsom:2018}: {longitudinal magnetic fields and average large-scale field strengths derived from the visible disc of the ZDI maps, $B_\mathrm{ZDI} = \sqrt{B_r^2+B_\theta^2+B_\phi^2}$}. 

We do not find a consistent relationship when comparing the total magnetic field variability with disc-integrated large-scale field strengths derived from ZDI maps at the corresponding rotational phases. Among our targets, one shows no correlation, two show a positive correlation, and one displays an anti-correlation (see Fig.~\ref{fig_extra:corr-small-large} {and the correlation coefficients therein}). This lack of a systematic trend suggests that magnetic field strengths obtained from ZDI may not reliably trace variations in stellar activity. This interpretation is supported by \citet{lehmann:2021}, who demonstrated that surface-averaged magnetic field strengths obtained from ZDI are not expected to reflect the evolution of the solar activity cycle, particularly when small-scale components dominate the total field.

When comparing the total magnetic field strength with the absolute value of $B_\ell$, we find a moderate anti-correlation for HIP 76768 and Mel25-5, with Spearman coefficients of $\rho = -0.6$ and $-0.58$, respectively. One interpretation of this result is that regions with the strongest magnetic field (higher $B_I$) also possess more complex and tangled field structures, leading to an enhanced cancellation of the polarisation signal and a decrease in $|B_\ell|$. Another possibility, {as signal cancellation can also be caused by well separated surface features}, could be that as more magnetically active regions come into view, increasing the small-scale field strength on the visible disc, the probability of these regions cancelling each other out increases. 

\section{Conclusion}
\label{sec:summary}
From this work, it is clear that we can detect rotational modulation of the disc-integrated total unsigned magnetic field on active stars. While this has been reported in the past \citep[e.g.][]{kochukhov:2017, donati:2023}, we have also shown, similar to \cite{cristofari:2025}, that the rotational period can be self-consistently recovered from the total unsigned magnetic field variation. Recovery of rotational periods has also been seen using other spectroscopic methods related to activity, such as starspot variability \citep{tang:2024} and equivalent widths of activity sensitive lines \citep{gomesdasilva:2025}. This apparent variability has some interesting consequences; it is already known that activity variation can influence the determination of accurate stellar parameters \citep[e.g.][]{spina:2020}. 
The short-term variations detected here could further complicate things, as accurate determination of stellar parameters would need to be performed simultaneously with an activity analysis to avoid any systematic effects from variability. 

When comparing the magnetic field with other activity indicators, we find coherence between the two. Given that there is no clear correlation between the activity and the large-scale field on rotational timescales (see Fig~\ref{fig_extra:corr-small-large}), it shows that the chromospheric emission is connected to unresolved magnetic regions on the stellar surface. This could be similar to reconnection due to sunspots that drive the chromospheric emission of the Sun. 
While activity indices appear to follow the magnetic field variation well on individual stars, there is a significant discrepancy between different stars. This means that activity indices appear to be well suited for monitoring activity evolution, but they are less reliable for studying activity relationships of different stars.

One clear limitation found in this work is the precision of magnetic field measurements in the context of studying variability. Even for stars with an age of a few hundred million years, the activity level is not sufficient to clearly detect rotational modulation. While studies in the near-infrared \citep[e.g.][]{hahlin:2025,cristofari:2025} have obtained higher magnetic field precision thanks to the lines available in that domain, we would still be unable to investigate stars with activity levels comparable to the Sun. Such work would require the development of more precise methods for the measurement of stellar magnetism, such as an application of the multiline technique described by \cite{lienhard:2023} on stars other than the Sun.

\begin{acknowledgements}
We thank the referee for their comments to help improve the paper. A.H. acknowledges support by the Swedish Research Council (projects 2019-03548 and 2023-03667). B.Z. acknowledges financial support from CAPES-PrInt (program $\#88887.939462/2024$-$00$). C.P.F. acknowledges funding from the European Union's Horizon Europe research and innovation programme under grant agreement No. 101079231 (EXOHOST), and from the United Kingdom Research and Innovation (UKRI) Horizon Europe Guarantee Scheme (grant number 10051045). K.A. acknowledges support from the Swiss National Science Foundation (SNSF) under the Postdoc Mobility grant P500PT\_230225. This research was supported by the Munich Institute for Astro-, Particle and BioPhysics (MIAPbP), which is funded by the Deutsche Forschungsgemeinschaft (DFG, German Research Foundation) under Germany´s Excellence Strategy – EXC-2094 – 390783311.

Based on observations obtained at the Canada–France– Hawaii Telescope (CFHT) which is operated from the summit of Maunakea by the National Research Council of Canada, the institut National des Sciences de l’Univers of the Centre National de la Recherche Scientifique of France, and the University of Hawaii. The observations at the Canada–France–Hawaii Telescope were performed with care and respect from the summit of Maunakea which is a significant cultural and historic site.
\end{acknowledgements}


\begin{thebibliography}{57}
\expandafter\ifx\csname natexlab\endcsname\relax\def\natexlab#1{#1}\fi

\bibitem[{{Amazo-G{\'o}mez} {et~al.}(2023){Amazo-G{\'o}mez}, {Alvarado-G{\'o}mez}, {Poppenh{\"a}ger}, {Hussain}, {Wood}, {Drake}, {do Nascimento}, {Anthony}, {Sanz-Forcada}, {Stelzer}, {Del Sordo}, {Damasso}, {Redfield}, {Donati}, {K{\"o}nig}, {H{\'e}brard}, \& {Miles-P{\'a}ez}}]{amazogomez:2023}
{Amazo-G{\'o}mez}, E.~M., {Alvarado-G{\'o}mez}, J.~D., {Poppenh{\"a}ger}, K., {et~al.} 2023, \mnras, 524, 5725

\bibitem[{{Anfinogentov} {et~al.}(2021){Anfinogentov}, {Nakariakov}, {Pascoe}, \& {Goddard}}]{anfinogentov:2021}
{Anfinogentov}, S.~A., {Nakariakov}, V.~M., {Pascoe}, D.~J., \& {Goddard}, C.~R. 2021, \apjs, 252, 11

\bibitem[{{Anthony} {et~al.}(2022){Anthony}, {N{\'u}{\~n}ez}, {Ag{\"u}eros}, {Curtis}, {do Nascimento}, {Machado}, {Mann}, {Newton}, {Rampalli}, {Thao}, \& {Wood}}]{Anthony:2022}
{Anthony}, F., {N{\'u}{\~n}ez}, A., {Ag{\"u}eros}, M.~A., {et~al.} 2022, \aj, 163, 257

\bibitem[{{Astudillo-Defru} {et~al.}(2017){Astudillo-Defru}, {Delfosse}, {Bonfils}, {Forveille}, {Lovis}, \& {Rameau}}]{Astudillo:2017}
{Astudillo-Defru}, N., {Delfosse}, X., {Bonfils}, X., {et~al.} 2017, \aap, 600, A13

\bibitem[{{Baliunas} {et~al.}(1995){Baliunas}, {Donahue}, {Soon}, {Horne}, {Frazer}, {Woodard-Eklund}, {Bradford}, {Rao}, {Wilson}, {Zhang}, {Bennett}, {Briggs}, {Carroll}, {Duncan}, {Figueroa}, {Lanning}, {Misch}, {Mueller}, {Noyes}, {Poppe}, {Porter}, {Robinson}, {Russell}, {Shelton}, {Soyumer}, {Vaughan}, \& {Whitney}}]{Baliunas:1995}
{Baliunas}, S.~L., {Donahue}, R.~A., {Soon}, W.~H., {et~al.} 1995, \apj, 438, 269

\bibitem[{{Bellotti} {et~al.}(2025{\natexlab{a}}){Bellotti}, {L{\"u}ftinger}, {Boro Saikia}, {Folsom}, {Petit}, {Morin}, {G{\"u}del}, {Donati}, \& {Alecian}}]{bellotti:2025a}
{Bellotti}, S., {L{\"u}ftinger}, T., {Boro Saikia}, S., {et~al.} 2025{\natexlab{a}}, \aap, 700, A282

\bibitem[{{Bellotti} {et~al.}(2023){Bellotti}, {Morin}, {Lehmann}, {Folsom}, {Hussain}, {Petit}, {Donati}, {Lavail}, {Carmona}, {Martioli}, {Romano Zaire}, {Alecian}, {Moutou}, {Fouqu{\'e}}, {Alencar}, {Artigau}, {Boisse}, {Bouchy}, {Cadieux}, {Cloutier}, {Cook}, {Delfosse}, {Doyon}, {H{\'e}brard}, {Kochukhov}, \& {Wade}}]{bellotti:2023}
{Bellotti}, S., {Morin}, J., {Lehmann}, L.~T., {et~al.} 2023, \aap, 676, A56

\bibitem[{{Bellotti} {et~al.}(2025{\natexlab{b}}){Bellotti}, {Petit}, {Jeffers}, {Marsden}, {Morin}, {Vidotto}, {Folsom}, {See}, \& {do Nascimento}}]{bellotti:2025}
{Bellotti}, S., {Petit}, P., {Jeffers}, S.~V., {et~al.} 2025{\natexlab{b}}, \aap, 693, A269

\bibitem[{{Boro Saikia} {et~al.}(2018){Boro Saikia}, {Lueftinger}, {Jeffers}, {Folsom}, {See}, {Petit}, {Marsden}, {Vidotto}, {Morin}, {Reiners}, {Guedel}, \& {BCool Collaboration}}]{boro-saikia:2018}
{Boro Saikia}, S., {Lueftinger}, T., {Jeffers}, S.~V., {et~al.} 2018, \aap, 620, L11

\bibitem[{{Boudreaux} {et~al.}(2022){Boudreaux}, {Newton}, {Mondrik}, {Charbonneau}, \& {Irwin}}]{Boudreaux:2022}
{Boudreaux}, E.~M., {Newton}, E.~R., {Mondrik}, N., {Charbonneau}, D., \& {Irwin}, J. 2022, \apj, 929, 80

\bibitem[{{Brown} {et~al.}(2022){Brown}, {Jeffers}, {Marsden}, {Morin}, {Boro Saikia}, {Petit}, {Jardine}, {See}, {Vidotto}, {Mengel}, {Dahlkemper}, \& {the BCool Collaboration}}]{brown:2022}
{Brown}, E.~L., {Jeffers}, S.~V., {Marsden}, S.~C., {et~al.} 2022, \mnras, 514, 4300

\bibitem[{{Cretignier} {et~al.}(2024){Cretignier}, {Pietrow}, \& {Aigrain}}]{Cretignier:2024}
{Cretignier}, M., {Pietrow}, A.~G.~M., \& {Aigrain}, S. 2024, \mnras, 527, 2940

\bibitem[{{Cristofari} {et~al.}(2025){Cristofari}, {Donati}, {Bellotti}, {Artigau}, {Carmona}, {Moutou}, {Delfosse}, {Petit}, {Finociety}, \& {Dias do Nascimento}}]{cristofari:2025}
{Cristofari}, P.~I., {Donati}, J.~F., {Bellotti}, S., {et~al.} 2025, \aap, 702, A111

\bibitem[{{Cristofari} {et~al.}(2023){Cristofari}, {Donati}, {Folsom}, {Masseron}, {Fouqu{\'e}}, {Moutou}, {Artigau}, {Carmona}, {Petit}, {Delfosse}, {Martioli}, \& {the SLS consortium}}]{cristofari:2023}
{Cristofari}, P.~I., {Donati}, J.~F., {Folsom}, C.~P., {et~al.} 2023, \mnras, 522, 1342

\bibitem[{{Delorme} {et~al.}(2011){Delorme}, {Collier Cameron}, {Hebb}, {Rostron}, {Lister}, {Norton}, {Pollacco}, \& {West}}]{delorme:2011}
{Delorme}, P., {Collier Cameron}, A., {Hebb}, L., {et~al.} 2011, \mnras, 413, 2218

\bibitem[{{Donati} {et~al.}(2023){Donati}, {Cristofari}, {Finociety}, {Klein}, {Moutou}, {Gaidos}, {Cadieux}, {Artigau}, {Correia}, {Bou{\'e}}, {Cook}, {Carmona}, {Lehmann}, {Bouvier}, {Martioli}, {Morin}, {Fouqu{\'e}}, {Delfosse}, {Doyon}, {H{\'e}brard}, {Alencar}, {Laskar}, {Arnold}, {Petit}, {K{\'o}sp{\'a}l}, {Vidotto}, {Folsom}, \& {collaboration}}]{donati:2023}
{Donati}, J.~F., {Cristofari}, P.~I., {Finociety}, B., {et~al.} 2023, \mnras, 525, 455

\bibitem[{{Donati} \& {Landstreet}(2009)}]{donati:2009}
{Donati}, J.~F. \& {Landstreet}, J.~D. 2009, \araa, 47, 333

\bibitem[{{Donati} {et~al.}(1997){Donati}, {Semel}, {Carter}, {Rees}, \& {Collier Cameron}}]{donati:1997}
{Donati}, J.~F., {Semel}, M., {Carter}, B.~D., {Rees}, D.~E., \& {Collier Cameron}, A. 1997, \mnras, 291, 658

\bibitem[{{Doyle} {et~al.}(2014){Doyle}, {Davies}, {Smalley}, {Chaplin}, \& {Elsworth}}]{doyle:2014}
{Doyle}, A.~P., {Davies}, G.~R., {Smalley}, B., {Chaplin}, W.~J., \& {Elsworth}, Y. 2014, \mnras, 444, 3592

\bibitem[{{Folsom} {et~al.}(2018){Folsom}, {Bouvier}, {Petit}, {L{\`e}bre}, {Amard}, {Palacios}, {Morin}, {Donati}, \& {Vidotto}}]{folsom:2018}
{Folsom}, C.~P., {Bouvier}, J., {Petit}, P., {et~al.} 2018, \mnras, 474, 4956

\bibitem[{{Folsom} {et~al.}(2016){Folsom}, {Petit}, {Bouvier}, {L{\`e}bre}, {Amard}, {Palacios}, {Morin}, {Donati}, {Jeffers}, {Marsden}, \& {Vidotto}}]{folsom:2016}
{Folsom}, C.~P., {Petit}, P., {Bouvier}, J., {et~al.} 2016, \mnras, 457, 580

\bibitem[{{Fritzewski} {et~al.}(2021){Fritzewski}, {Barnes}, {James}, {J{\"a}rvinen}, \& {Strassmeier}}]{Fritzewski:2021}
{Fritzewski}, D.~J., {Barnes}, S.~A., {James}, D.~J., {J{\"a}rvinen}, S.~P., \& {Strassmeier}, K.~G. 2021, \aap, 656, A103

\bibitem[{{Gomes da Silva} {et~al.}(2025){Gomes da Silva}, {Delgado-Mena}, {Santos}, {Monteiro}, {Larue}, {Su{\'a}rez Mascare{\~n}o}, {Delfosse}, {Mignon}, {Artigau}, {Nari}, {Abreu}, {Aguiar}, {Al Moulla}, {Allain}, {Allart}, {Arial}, {Auger}, {Baron}, {Barros}, {Bazinet}, {Benneke}, {Blind}, {Bohlender}, {Boisse}, {Bonfils}, {Boucher}, {Bouchy}, {Bourrier}, {Bovay}, {Branco}, {Broeg}, {Brousseau}, {Bruniquel}, {Bryan}, {Cabral}, {Cadieux}, {Canto Martins}, {Carmona}, {Carteret}, {Challita}, {Chazelas}, {Cloutier}, {Coelho}, {Cointepas}, {Conod}, {Cook}, {Costa Silva}, {Cowan}, {Cristo}, {Darveau-Bernier}, {Dauplaise}, {de Lima Gomes}, {De Medeiros}, {Delisle}, {Doshi}, {Doyon}, {Dumusque}, {Ehrenreich}, {Figueira}, {Fontinele}, {Forveille}, {Frensch}, {Gagn{\'e}}, {Genest}, {Genolet}, {Gonz{\'a}lez Hern{\'a}ndez}, {Glover}, {Gracia T{\'e}mich}, {Grieves}, {Gromek}, {Hernandez}, {Hobson}, {Hoeijmakers}, {Hubin}, {Jahandar}, {Jayawardhana}, {K{\"a}ufl}, {Kerley}, {Kolb}, {Krishnamurthy}, {Kung}, {L'Heureux},
  {Lafreni{\`e}re}, {Lamontagne}, {de Castro Le{\~a}o}, {Leath}, {Lim}, {Lipper}, {Lo Curto}, {Lovis}, {Malo}, {Martins}, {Matthews}, {Mayer}, {Melo}, {Messamah}, {Messias}, {Metchev}, {Moranta}, {Mordasini}, {Mounzer}, {Mraz}, {Nielsen}, {Osborn}, {Otegi}, {Ouellet}, {Parc}, {Pasquini}, {Passegger}, {Pelletier}, {Pepe}, {Peroux}, {Piaulet-Ghorayeb}, {Plotnykov}, {Pompei}, {Poulin-Girard}, {Rasilla}, {Rebolo}, {Reshetov}, {Rowe}, {Saint-Antoine}, {Sarajlic}, {Saviane}, {Schnell}, {Segovia}, {S{\'e}gransan}, {Seidel}, {Silber}, {Sinclair}, {Sordet}, {Sosnowska}, {Srivastava}, {Stefanov}, {Teixeira}, {Thibault}, {Udry}, {Valencia}, {Vall{\'e}e}, {Vandal}, {Vaulato}, {Wade}, {Wardenier}, {Wehbe}, {Weisserman}, {Wevers}, {Wildi}, {Yariv}, \& {Zins}}]{gomesdasilva:2025}
{Gomes da Silva}, J., {Delgado-Mena}, E., {Santos}, N.~C., {et~al.} 2025, \aap, 700, A177

\bibitem[{{Gomes da Silva} {et~al.}(2021){Gomes da Silva}, {Santos}, {Adibekyan}, {Sousa}, {Campante}, {Figueira}, {Bossini}, {Delgado-Mena}, {Monteiro}, {de Laverny}, {Recio-Blanco}, \& {Lovis}}]{Gomes:2021}
{Gomes da Silva}, J., {Santos}, N.~C., {Adibekyan}, V., {et~al.} 2021, \aap, 646, A77

\bibitem[{{Gustafsson} {et~al.}(2008){Gustafsson}, {Edvardsson}, {Eriksson}, {J{\o}rgensen}, {Nordlund}, \& {Plez}}]{gustafsson:2008}
{Gustafsson}, B., {Edvardsson}, B., {Eriksson}, K., {et~al.} 2008, \aap, 486, 951

\bibitem[{{Hahlin} \& {Kochukhov}(2022)}]{hahlin:2022}
{Hahlin}, A. \& {Kochukhov}, O. 2022, \aap, 659, A151

\bibitem[{{Hahlin} {et~al.}(2025){Hahlin}, {Kochukhov}, {Chaturvedi}, {Guenther}, {Hatzes}, {Heiter}, {Lavail}, {Nagel}, {Piskunov}, {Pouilly}, {Rains}, {Reiners}, {Rengel}, {Seeman}, \& {Shulyak}}]{hahlin:2025}
{Hahlin}, A., {Kochukhov}, O., {Chaturvedi}, P., {et~al.} 2025, \aap, 696, A4

\bibitem[{{Hahlin} {et~al.}(2023){Hahlin}, {Kochukhov}, {Rains}, {Lavail}, {Hatzes}, {Piskunov}, {Reiners}, {Seemann}, {Boldt-Christmas}, {Guenther}, {Heiter}, {Nortmann}, {Yan}, {Shulyak}, {Smoker}, {Rodler}, {Bristow}, {Dorn}, {Jung}, {Marquart}, \& {Stempels}}]{hahlin:2023}
{Hahlin}, A., {Kochukhov}, O., {Rains}, A.~D., {et~al.} 2023, \aap, 675, A91

\bibitem[{{Haywood} {et~al.}(2016){Haywood}, {Collier Cameron}, {Unruh}, {Lovis}, {Lanza}, {Llama}, {Deleuil}, {Fares}, {Gillon}, {Moutou}, {Pepe}, {Pollacco}, {Queloz}, \& {S{\'e}gransan}}]{haywood:2016}
{Haywood}, R.~D., {Collier Cameron}, A., {Unruh}, Y.~C., {et~al.} 2016, \mnras, 457, 3637

\bibitem[{{Kochukhov}(2016)}]{kochukhov:2016}
{Kochukhov}, O. 2016, {Doppler and Zeeman Doppler Imaging of Stars}, ed. J.-P. {Rozelot} \& C.~{Neiner}, Vol. 914, 177

\bibitem[{{Kochukhov}(2021)}]{kochukhov:2021}
{Kochukhov}, O. 2021, \aapr, 29, 1

\bibitem[{{Kochukhov} {et~al.}(2020){Kochukhov}, {Hackman}, {Lehtinen}, \& {Wehrhahn}}]{kochukhov:2020a}
{Kochukhov}, O., {Hackman}, T., {Lehtinen}, J.~J., \& {Wehrhahn}, A. 2020, \aap, 635, A142

\bibitem[{{Kochukhov} \& {Lavail}(2017)}]{kochukhov:2017}
{Kochukhov}, O. \& {Lavail}, A. 2017, \apjl, 835, L4

\bibitem[{{Kochukhov}(2007)}]{kochukhov:2007}
{Kochukhov}, O.~P. 2007, in Physics of Magnetic Stars, ed. I.~I. {Romanyuk}, D.~O. {Kudryavtsev}, O.~M. {Neizvestnaya}, \& V.~M. {Shapoval} (Special Astrophysical Observatory of the Russian Academy of Science,), 109--118

\bibitem[{{Lavail} {et~al.}(2019){Lavail}, {Kochukhov}, \& {Hussain}}]{lavail:2019}
{Lavail}, A., {Kochukhov}, O., \& {Hussain}, G.~A.~J. 2019, \aap, 630, A99

\bibitem[{{Lehmann} {et~al.}(2021){Lehmann}, {Hussain}, {Vidotto}, {Jardine}, \& {Mackay}}]{lehmann:2021}
{Lehmann}, L.~T., {Hussain}, G.~A.~J., {Vidotto}, A.~A., {Jardine}, M.~M., \& {Mackay}, D.~H. 2021, \mnras, 500, 1243

\bibitem[{{Lehmann} {et~al.}(2015){Lehmann}, {K{\"u}nstler}, {Carroll}, \& {Strassmeier}}]{lehmann:2015}
{Lehmann}, L.~T., {K{\"u}nstler}, A., {Carroll}, T.~A., \& {Strassmeier}, K.~G. 2015, Astronomische Nachrichten, 336, 258

\bibitem[{{Lienhard} {et~al.}(2023){Lienhard}, {Mortier}, {Cegla}, {Cameron}, {Klein}, \& {Watson}}]{lienhard:2023}
{Lienhard}, F., {Mortier}, A., {Cegla}, H.~M., {et~al.} 2023, \mnras, 522, 5862

\bibitem[{{Marsden} {et~al.}(2014){Marsden}, {Petit}, {Jeffers}, {Morin}, {Fares}, {Reiners}, {do Nascimento}, {Auri{\`e}re}, {Bouvier}, {Carter}, {Catala}, {Dintrans}, {Donati}, {Gastine}, {Jardine}, {Konstantinova-Antova}, {Lanoux}, {Ligni{\`e}res}, {Morgenthaler}, {Ram{\`\i}rez-V{\`e}lez}, {Th{\'e}ado}, {Van Grootel}, \& {BCool Collaboration}}]{marsden:2014}
{Marsden}, S.~C., {Petit}, P., {Jeffers}, S.~V., {et~al.} 2014, \mnras, 444, 3517

\bibitem[{{Messina} {et~al.}(2010){Messina}, {Desidera}, {Turatto}, {Lanzafame}, \& {Guinan}}]{messina:2010}
{Messina}, S., {Desidera}, S., {Turatto}, M., {Lanzafame}, A.~C., \& {Guinan}, E.~F. 2010, 520, A15

\bibitem[{{Messina} \& {Guinan}(2002)}]{Messina:2002}
{Messina}, S. \& {Guinan}, E.~F. 2002, \aap, 393, 225

\bibitem[{{Newton} {et~al.}(2017){Newton}, {Irwin}, {Charbonneau}, {Berlind}, {Calkins}, \& {Mink}}]{Newton:2017}
{Newton}, E.~R., {Irwin}, J., {Charbonneau}, D., {et~al.} 2017, \apj, 834, 85

\bibitem[{{Petit} {et~al.}(2021){Petit}, {Folsom}, {Donati}, {Yu}, {do Nascimento}, {Jeffers}, {Marsden}, {Morin}, \& {Vidotto}}]{petit:2021}
{Petit}, P., {Folsom}, C.~P., {Donati}, J.~F., {et~al.} 2021, \aap, 648, A55

\bibitem[{{Pouilly} {et~al.}(2024){Pouilly}, {Audard}, {K{\'o}sp{\'a}l}, \& {Lavail}}]{pouilly:2024}
{Pouilly}, K., {Audard}, M., {K{\'o}sp{\'a}l}, {\'A}., \& {Lavail}, A. 2024, \aap, 691, A18

\bibitem[{{Reiners} {et~al.}(2022){Reiners}, {Shulyak}, {K{\"a}pyl{\"a}}, {Ribas}, {Nagel}, {Zechmeister}, {Caballero}, {Shan}, {Fuhrmeister}, {Quirrenbach}, {Amado}, {Montes}, {Jeffers}, {Azzaro}, {B{\'e}jar}, {Chaturvedi}, {Henning}, {K{\"u}rster}, \& {Pall{\'e}}}]{Reiners:2022}
{Reiners}, A., {Shulyak}, D., {K{\"a}pyl{\"a}}, P.~J., {et~al.} 2022, \aap, 662, A41

\bibitem[{{Rescigno} {et~al.}(2024){Rescigno}, {Mortier}, {Dumusque}, {Lakeland}, {Haywood}, {Piskunov}, {Nicholson}, {L{\'o}pez-Morales}, {Dalal}, {Cretignier}, {Klein}, {Cameron}, {Ghedina}, {Gonzalez}, {Cosentino}, {Sozzetti}, \& {Saar}}]{rescigno:2024}
{Rescigno}, F., {Mortier}, A., {Dumusque}, X., {et~al.} 2024, \mnras, 532, 2741

\bibitem[{{Robinson}(1980)}]{robinson:1980}
{Robinson}, Jr., R.~D. 1980, \apj, 239, 961

\bibitem[{{Ryabchikova} {et~al.}(2015){Ryabchikova}, {Piskunov}, {Kurucz}, {Stempels}, {Heiter}, {Pakhomov}, \& {Barklem}}]{ryabchikova:2015}
{Ryabchikova}, T., {Piskunov}, N., {Kurucz}, R.~L., {et~al.} 2015, \physscr, 90, 054005

\bibitem[{{See} {et~al.}(2016){See}, {Jardine}, {Vidotto}, {Donati}, {Boro Saikia}, {Bouvier}, {Fares}, {Folsom}, {Gregory}, {Hussain}, {Jeffers}, {Marsden}, {Morin}, {Moutou}, {do Nascimento}, {Petit}, \& {Waite}}]{see:2016}
{See}, V., {Jardine}, M., {Vidotto}, A.~A., {et~al.} 2016, \mnras, 462, 4442

\bibitem[{{Sharma}(2017)}]{sharma:2017}
{Sharma}, S. 2017, \araa, 55, 213

\bibitem[{{Shulyak} {et~al.}(2019){Shulyak}, {Reiners}, {Nagel}, {Tal-Or}, {Caballero}, {Zechmeister}, {B{\'e}jar}, {Cort{\'e}s-Contreras}, {Martin}, {Kaminski}, {Ribas}, {Quirrenbach}, {Amado}, {Anglada-Escud{\'e}}, {Bauer}, {Dreizler}, {Guenther}, {Henning}, {Jeffers}, {K{\"u}rster}, {Lafarga}, {Montes}, {Morales}, \& {Pedraz}}]{shulyak:2019}
{Shulyak}, D., {Reiners}, A., {Nagel}, E., {et~al.} 2019, \aap, 626, A86

\bibitem[{{Smette} {et~al.}(2015){Smette}, {Sana}, {Noll}, {Horst}, {Kausch}, {Kimeswenger}, {Barden}, {Szyszka}, {Jones}, {Gallenne}, {Vinther}, {Ballester}, \& {Taylor}}]{smette:2015}
{Smette}, A., {Sana}, H., {Noll}, S., {et~al.} 2015, \aap, 576, A77

\bibitem[{{Spina} {et~al.}(2020){Spina}, {Nordlander}, {Casey}, {Bedell}, {D'Orazi}, {Mel{\'e}ndez}, {Karakas}, {Desidera}, {Baratella}, {Yana Galarza}, \& {Casali}}]{spina:2020}
{Spina}, L., {Nordlander}, T., {Casey}, A.~R., {et~al.} 2020, \apj, 895, 52

\bibitem[{{Tang} {et~al.}(2024){Tang}, {Johns-Krull}, {Prato}, \& {Stahl}}]{tang:2024}
{Tang}, S.-Y., {Johns-Krull}, C.~M., {Prato}, L., \& {Stahl}, A.~G. 2024, \apj, 973, 124

\bibitem[{{Vidotto} {et~al.}(2014){Vidotto}, {Gregory}, {Jardine}, {Donati}, {Petit}, {Morin}, {Folsom}, {Bouvier}, {Cameron}, {Hussain}, {Marsden}, {Waite}, {Fares}, {Jeffers}, \& {do Nascimento}}]{vidotto:2014}
{Vidotto}, A.~A., {Gregory}, S.~G., {Jardine}, M., {et~al.} 2014, \mnras, 441, 2361

\bibitem[{{Yamashita} {et~al.}(2025){Yamashita}, {Itoh}, \& {Toriumi}}]{yamashita:2025}
{Yamashita}, M., {Itoh}, Y., \& {Toriumi}, S. 2025, \apj, 985, 46

\bibitem[{{Zechmeister} \& {K{\"u}rster}(2009)}]{zechmeister:2009}
{Zechmeister}, M. \& {K{\"u}rster}, M. 2009, \aap, 496, 577

\end{thebibliography}

\onecolumn
\begin{appendix}
\section{Magnetic variability}
\label{app:periodograms}
\begin{figure*}[h]
    \centering
    \includegraphics[width=0.45\linewidth]{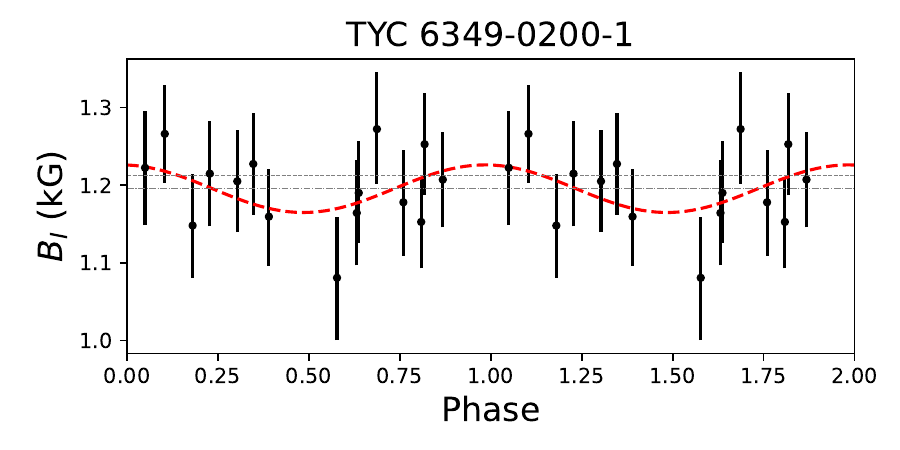}
    \includegraphics[width=0.45\linewidth]{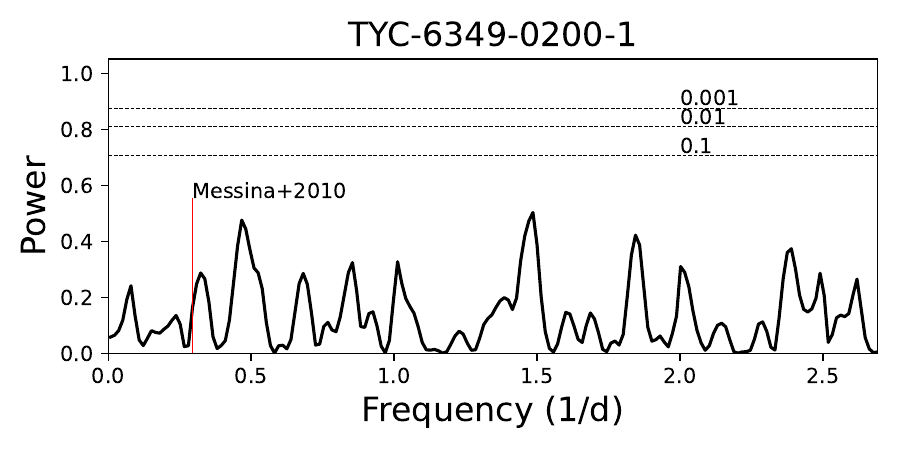}
    \includegraphics[width=0.45\linewidth]{Figures/HIP_Variability_phasefolded.pdf}
    \includegraphics[width=0.45\linewidth]{Figures/HIP_Periodogram.pdf}
    \includegraphics[width=0.45\linewidth]{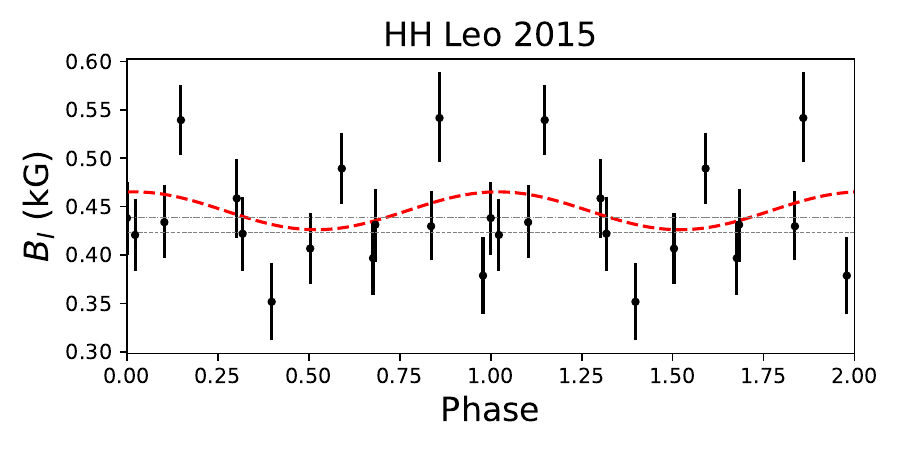}
    \includegraphics[width=0.45\linewidth]{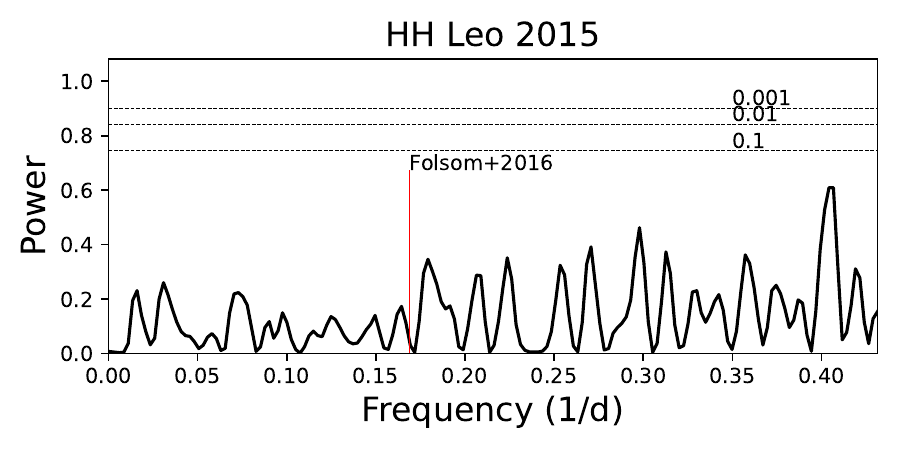}
    \includegraphics[width=0.45\linewidth]{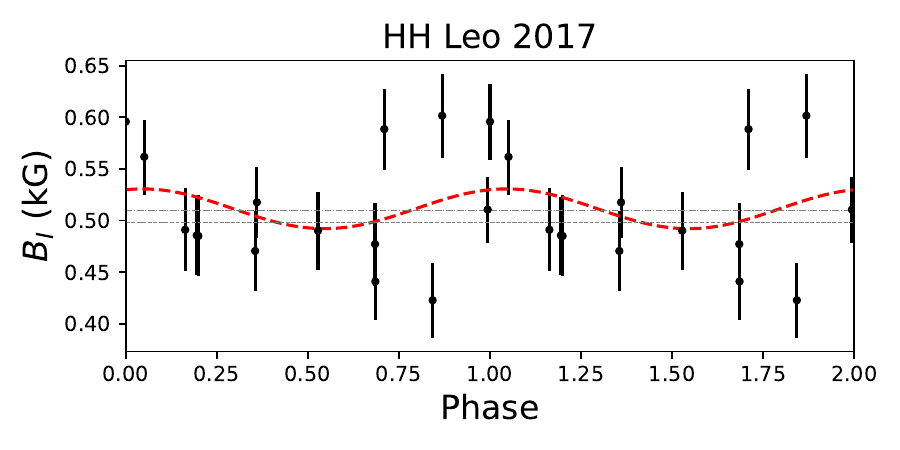}
    \includegraphics[width=0.45\linewidth]{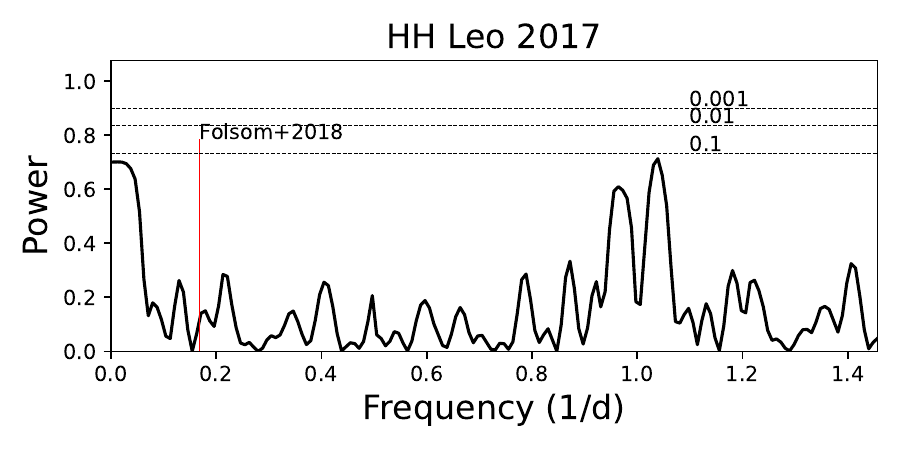}
    \includegraphics[width=0.45\linewidth]{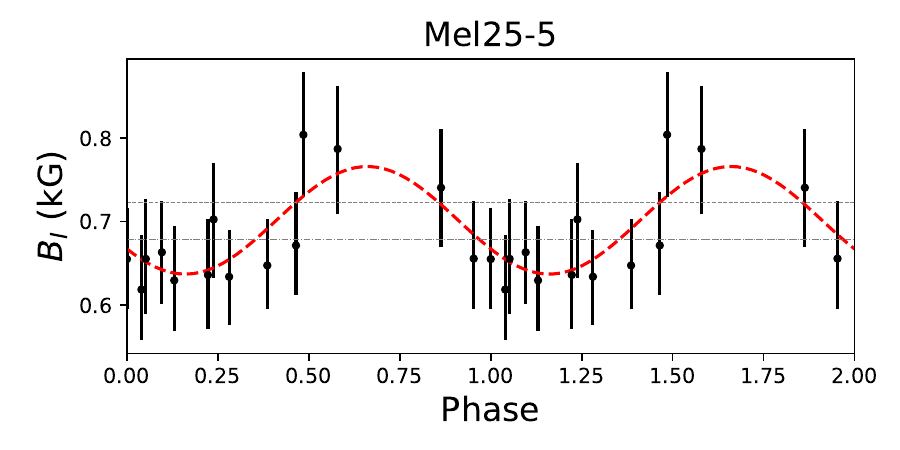}
    \includegraphics[width=0.45\linewidth]{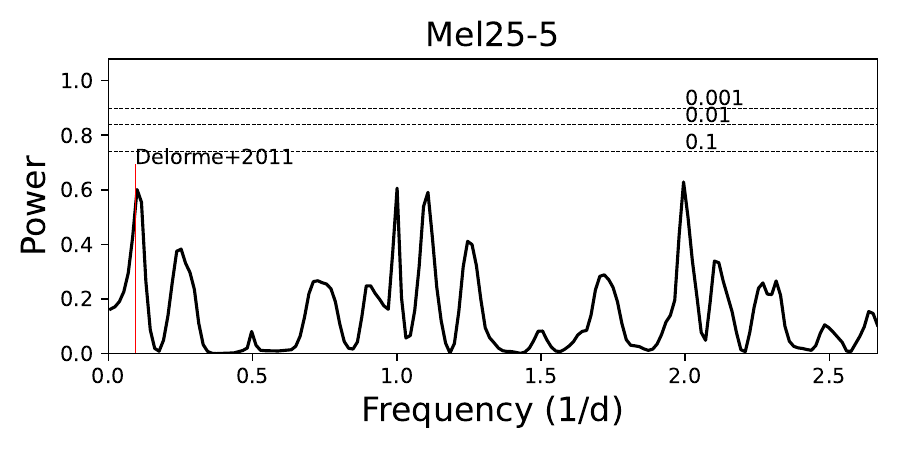}
    \caption{Magnetic variability for the stars studied in this work. Left column: Phase-folded magnetic field data along with the best fit sinusoidal function with periods from Table~\ref{tab:stellar-parameters}. The average magnetic field, both from the averaged spectra and the sinusoidal fit are shown as dotted and dashed lines, respectively. Right column: Periodograms of the magnetic sample, red lines marks the literature periods used for the sinusoidal fitting. False alarm probabilities are marked with dashed lines.}
    \label{fig:magnetic_variability}
\end{figure*}
\newpage
\section{Activity correlations}

\begin{figure*}
    \centering
    \includegraphics[width=\linewidth]{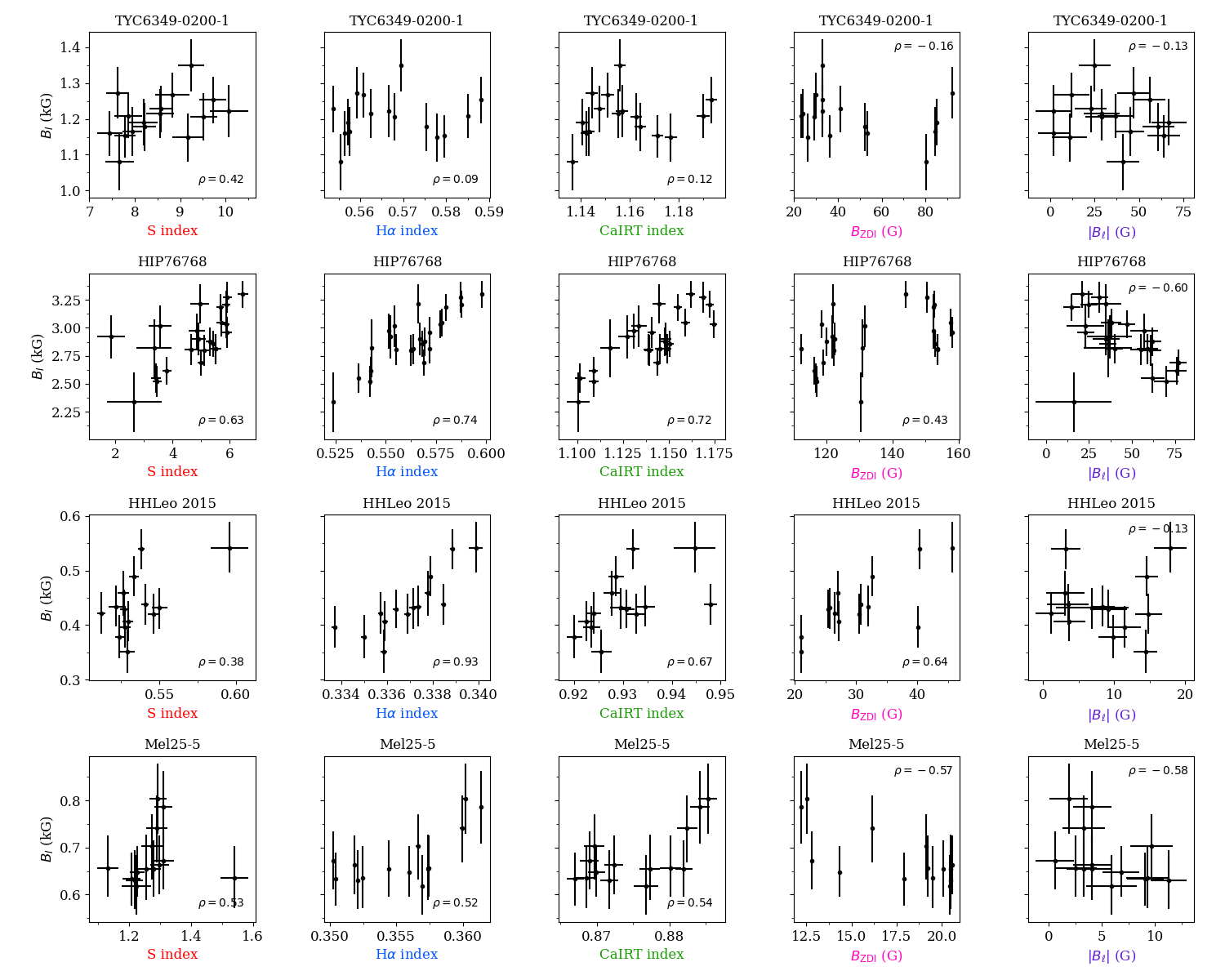}[h]
    \caption{Correlations between the {total unsigned} magnetic field, activity traces and other large-scale polarimetric field
measurements. The Spearman correlation coefficient is shown in each panel.}
    \label{fig_extra:corr-small-large}
\end{figure*}

\begin{figure}[h]
    \centering
    \includegraphics[width=0.5\linewidth]{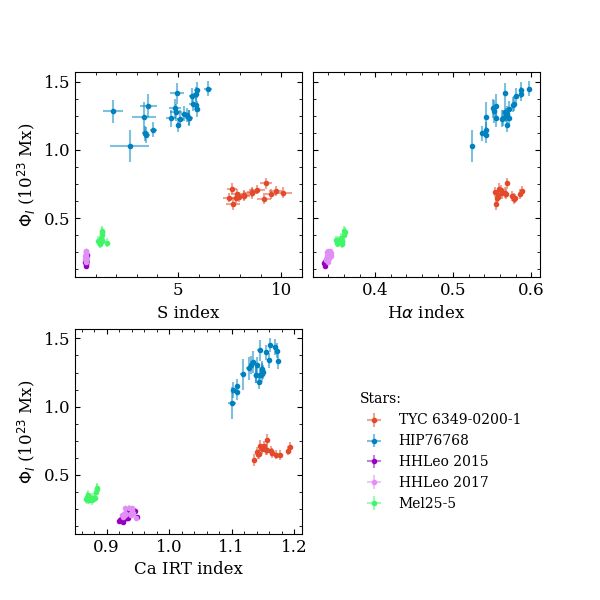}
    \caption{Similar to Fig.~\ref{fig:activity-correlation}, but for correlations between the unsigned magnetic flux and other activity indicators.}
    \label{fig_extra:activity-correlationFlux}
\end{figure}

\newpage
\section{Individual measurements}

{\small
\begin{longtable}[c]{lrrrrrrrrr}
    
    \caption{Magnetic field and activity data for individual measurements}
    \label{tab:activity_data}\\
    
        \hline\hline
        Star & MJD-24.5e5 & $B_I$\,(kG) & $B_V$\,(G) & $B_z$\,(G) & S index & H$\alpha$ index & CaIRT index & S/N & $\chi2_{\rm reduced}^{*}$ \\
        \hline
        HIP 76768$^{1}$& 6430.872&3.30$\pm$0.12& 144&-21.0$\pm$6.0& 6.45$\pm$0.18 & 0.5978$\pm$0.0004 & 1.162$\pm$0.002 &107&2.99\\
 & 6431.070&3.28$\pm$0.14& 150&-31.0$\pm$5.0& 5.91$\pm$0.16 & 0.5873$\pm$0.0003 & 1.169$\pm$0.002 &122&4.49\\
 & 6431.828&3.05$\pm$0.12& 158&-38.0$\pm$6.0& 5.70$\pm$0.17 & 0.5780$\pm$0.0004 & 1.159$\pm$0.002 &110&3.50\\
 & 6432.920&2.80$\pm$0.14& 122&-61.0$\pm$6.0& 5.11$\pm$0.18 & 0.5624$\pm$0.0004 & 1.139$\pm$0.003 &101&3.52\\
 & 6433.066&2.88$\pm$0.13& 120&-62.0$\pm$5.0& 5.30$\pm$0.14 & 0.5694$\pm$0.0003 & 1.148$\pm$0.002 &124&3.94\\
 & 6433.786&2.81$\pm$0.13& 112&-59.0$\pm$6.0& 5.51$\pm$0.20 & 0.5635$\pm$0.0004 & 1.145$\pm$0.002 &108&3.71\\
 & 6434.014&3.03$\pm$0.13& 119&-47.0$\pm$5.0& 5.87$\pm$0.13 & 0.5772$\pm$0.0003 & 1.175$\pm$0.002 &128&4.39\\
 & 6434.898&3.18$\pm$0.12& 152&-15.0$\pm$5.0& 5.66$\pm$0.13 & 0.5801$\pm$0.0003 & 1.155$\pm$0.002 &121&3.71\\
 & 6435.042&3.21$\pm$0.12& 153&-25.0$\pm$5.0& 5.87$\pm$0.14 & 0.5876$\pm$0.0003 & 1.172$\pm$0.002 &125&3.97\\
 & 6435.845&2.81$\pm$0.13& 154&-40.0$\pm$5.0& 5.50$\pm$0.12 & 0.5716$\pm$0.0003 & 1.149$\pm$0.002 &127&4.71\\
 & 6435.996&2.86$\pm$0.12& 153&-36.0$\pm$5.0& 5.41$\pm$0.11 & 0.5681$\pm$0.0003 & 1.151$\pm$0.002 &128&3.79\\
 & 6436.872&2.69$\pm$0.12& 119&-77.0$\pm$5.0& 5.00$\pm$0.11 & 0.5691$\pm$0.0003 & 1.144$\pm$0.002 &121&3.82\\
 & 6437.016&2.52$\pm$0.14& 117&-70.0$\pm$7.0& 3.46$\pm$0.16 & 0.5422$\pm$0.0004 & 1.109$\pm$0.003 &97&3.27\\
 & 6437.027&2.55$\pm$0.13& 117&-62.0$\pm$7.0& 3.41$\pm$0.17 & 0.5364$\pm$0.0004 & 1.101$\pm$0.003 &94&3.36\\
 & 6437.039&2.62$\pm$0.13& 116&-76.0$\pm$6.0& 3.80$\pm$0.16 & 0.5427$\pm$0.0004 & 1.109$\pm$0.003 &103&3.29\\
 & 6437.808&2.92$\pm$0.19& 122&-37.0$\pm$13.0& 1.84$\pm$0.49 & 0.5523$\pm$0.0007 & 1.127$\pm$0.005 &57&2.00\\
 & 6437.819&3.22$\pm$0.17& 122&-35.0$\pm$9.0& 4.95$\pm$0.32 & 0.5661$\pm$0.0005 & 1.145$\pm$0.004 &74&3.00\\
 & 6437.831&2.90$\pm$0.14& 123&-35.0$\pm$8.0& 4.90$\pm$0.27 & 0.5669$\pm$0.0005 & 1.148$\pm$0.003 &86&2.60\\
 & 6437.984&2.34$\pm$0.27& 130&-16.0$\pm$22.0& 2.65$\pm$0.95 & 0.5240$\pm$0.0009 & 1.100$\pm$0.006 &40&2.28\\
 & 6437.996&2.82$\pm$0.26& 131&-36.0$\pm$14.0& 3.35$\pm$0.61 & 0.5427$\pm$0.0008 & 1.118$\pm$0.005 &50&3.29\\
 & 6438.008&3.02$\pm$0.19& 132&-23.0$\pm$11.0& 3.56$\pm$0.40 & 0.5544$\pm$0.0006 & 1.134$\pm$0.004 &62&2.40\\
 & 6439.848&2.97$\pm$0.15& 153&-57.0$\pm$8.0& 4.85$\pm$0.29 & 0.5516$\pm$0.0005 & 1.131$\pm$0.003 &81&2.89\\
 & 6440.032&2.81$\pm$0.14& 154&-55.0$\pm$6.0& 4.66$\pm$0.23 & 0.5550$\pm$0.0004 & 1.139$\pm$0.003 &101&3.72\\
 & 6442.806&2.96$\pm$0.14& 158&-23.0$\pm$5.0& 5.91$\pm$0.16 & 0.5716$\pm$0.0003 & 1.141$\pm$0.002 &118&4.41\\
\hline
TYC6349-0200-1$^{1}$& 6458.432&1.35$\pm$0.07& 33&-25.0$\pm$9.0& 9.24$\pm$0.29 & 0.5694$\pm$0.0004 & 1.156$\pm$0.002 &117&5.94\\
 & 6459.465&1.21$\pm$0.07& 29&29.0$\pm$9.0& 9.51$\pm$0.30 & 0.5681$\pm$0.0004 & 1.163$\pm$0.002 &114&4.88\\
 & 6459.614&1.23$\pm$0.07& 41&23.0$\pm$9.0& 8.58$\pm$0.26 & 0.5538$\pm$0.0003 & 1.148$\pm$0.002 &115&3.98\\
 & 6460.400&1.08$\pm$0.08& 80&-41.0$\pm$9.0& 7.66$\pm$0.32 & 0.5555$\pm$0.0004 & 1.136$\pm$0.002 &112&7.14\\
 & 6460.603&1.19$\pm$0.07& 85&-67.0$\pm$10.0& 8.19$\pm$0.31 & 0.5572$\pm$0.0004 & 1.141$\pm$0.003 &104&3.07\\
 & 6462.456&1.15$\pm$0.07& 26&11.0$\pm$10.0& 9.18$\pm$0.35 & 0.5778$\pm$0.0004 & 1.177$\pm$0.003 &107&3.99\\
 & 6462.617&1.21$\pm$0.07& 24&29.0$\pm$10.0& 8.56$\pm$0.31 & 0.5625$\pm$0.0004 & 1.155$\pm$0.003 &106&4.21\\
 & 6464.432&1.18$\pm$0.07& 52&-61.0$\pm$9.0& 8.21$\pm$0.25 & 0.5753$\pm$0.0004 & 1.164$\pm$0.002 &117&5.59\\
 & 6464.599&1.15$\pm$0.06& 36&-64.0$\pm$9.0& 7.78$\pm$0.24 & 0.5795$\pm$0.0004 & 1.171$\pm$0.002 &114&3.12\\
 & 6465.420&1.22$\pm$0.07& 33&-2.0$\pm$10.0& 10.08$\pm$0.42 & 0.5667$\pm$0.0004 & 1.157$\pm$0.002 &107&5.26\\
 & 6465.605&1.27$\pm$0.06& 30&-12.0$\pm$10.0& 8.83$\pm$0.38 & 0.5607$\pm$0.0004 & 1.151$\pm$0.003 &101&3.34\\
 & 6467.405&1.16$\pm$0.07& 84&-45.0$\pm$8.0& 7.95$\pm$0.22 & 0.5576$\pm$0.0003 & 1.143$\pm$0.002 &120&4.90\\
 & 6467.594&1.27$\pm$0.07& 92&-47.0$\pm$9.0& 7.62$\pm$0.26 & 0.5592$\pm$0.0004 & 1.145$\pm$0.002 &109&4.35\\
 & 6471.451&1.25$\pm$0.07& 33&-56.0$\pm$9.0& 9.72$\pm$0.30 & 0.5881$\pm$0.0004 & 1.193$\pm$0.002 &113&3.49\\
 & 6471.621&1.21$\pm$0.06& 23&-37.0$\pm$10.0& 7.86$\pm$0.31 & 0.5851$\pm$0.0004 & 1.190$\pm$0.003 &104&2.55\\
 & 6473.401&1.16$\pm$0.06& 54&2.0$\pm$9.0& 7.45$\pm$0.28 & 0.5564$\pm$0.0004 & 1.142$\pm$0.002 &113&3.26\\
\hline
MEL 25-5$^{2}$& 7284.482&0.66$\pm$0.06& 20&3.3$\pm$1.6& 1.28$\pm$0.03 & 0.3544$\pm$0.0001 & 0.882$\pm$0.001 &199&14.49\\
 & 7285.497&0.66$\pm$0.06& 21&4.1$\pm$1.8& 1.30$\pm$0.03 & 0.3518$\pm$0.0001 & 0.872$\pm$0.001 &178&9.70\\
 & 7287.451&0.63$\pm$0.06& 18&9.1$\pm$1.7& 1.21$\pm$0.03 & 0.3504$\pm$0.0001 & 0.867$\pm$0.001 &191&10.29\\
 & 7288.560&0.65$\pm$0.05& 14&6.8$\pm$1.7& 1.23$\pm$0.02 & 0.3560$\pm$0.0001 & 0.870$\pm$0.001 &184&6.43\\
 & 7289.393&0.67$\pm$0.06& 13&0.6$\pm$1.8& 1.31$\pm$0.03 & 0.3503$\pm$0.0001 & 0.869$\pm$0.001 &179&7.58\\
 & 7289.608&0.80$\pm$0.07& 13&1.9$\pm$1.8& 1.29$\pm$0.03 & 0.3602$\pm$0.0001 & 0.885$\pm$0.001 &175&9.73\\
 & 7290.604&0.79$\pm$0.08& 12&-4.1$\pm$1.8& 1.31$\pm$0.03 & 0.3614$\pm$0.0001 & 0.884$\pm$0.001 &174&8.54\\
 & 7293.608&0.74$\pm$0.07& 16&3.3$\pm$2.0& 1.29$\pm$0.03 & 0.3599$\pm$0.0001 & 0.882$\pm$0.001 &166&6.94\\
 & 7294.557&0.66$\pm$0.06& 19&2.5$\pm$2.0& 1.13$\pm$0.03 & 0.3574$\pm$0.0001 & 0.880$\pm$0.001 &157&5.08\\
 & 7295.473&0.62$\pm$0.06& 20&5.9$\pm$2.4& 1.22$\pm$0.05 & 0.3570$\pm$0.0002 & 0.877$\pm$0.002 &137&4.54\\
 & 7295.601&0.66$\pm$0.07& 21&4.1$\pm$1.9& 1.25$\pm$0.03 & 0.3574$\pm$0.0001 & 0.877$\pm$0.001 &167&6.90\\
 & 7296.427&0.63$\pm$0.06& 21&11.3$\pm$1.7& 1.22$\pm$0.03 & 0.3521$\pm$0.0001 & 0.872$\pm$0.001 &189&7.42\\
 & 7297.406&0.64$\pm$0.07& 19&9.3$\pm$2.0& 1.54$\pm$0.04 & 0.3525$\pm$0.0001 & 0.869$\pm$0.001 &166&6.53\\
 & 7297.570&0.70$\pm$0.07& 19&9.7$\pm$2.0& 1.27$\pm$0.03 & 0.3566$\pm$0.0001 & 0.870$\pm$0.001 &163&6.60\\
\hline
HH Leo (2015)$^{2}$& 7088.497&0.44$\pm$0.04& 31&3.5$\pm$2.9& 0.54$\pm$0.00 & 0.3385$\pm$0.0001 & 0.948$\pm$0.001 &343&17.56\\
 & 7092.535&0.43$\pm$0.04& 26&-6.9$\pm$5.1& 0.55$\pm$0.01 & 0.3371$\pm$0.0002 & 0.930$\pm$0.002 &210&6.46\\
 & 7093.579&0.54$\pm$0.05& 46&17.9$\pm$2.3& 0.60$\pm$0.01 & 0.3399$\pm$0.0003 & 0.945$\pm$0.004 &120&2.75\\
 & 7094.544&0.42$\pm$0.04& 31&14.8$\pm$1.9& 0.55$\pm$0.00 & 0.3369$\pm$0.0001 & 0.933$\pm$0.002 &255&9.17\\
 & 7099.358&0.43$\pm$0.04& 25&9.2$\pm$2.6& 0.53$\pm$0.00 & 0.3364$\pm$0.0001 & 0.931$\pm$0.002 &311&12.42\\
 & 7114.510&0.35$\pm$0.04& 21&-14.4$\pm$1.7& 0.53$\pm$0.00 & 0.3359$\pm$0.0002 & 0.926$\pm$0.002 &225&7.86\\
 & 7133.396&0.49$\pm$0.04& 33&-14.6$\pm$1.6& 0.53$\pm$0.00 & 0.3379$\pm$0.0001 & 0.929$\pm$0.002 &300&12.83\\
 & 7136.425&0.43$\pm$0.04& 32&8.4$\pm$1.7& 0.52$\pm$0.00 & 0.3373$\pm$0.0001 & 0.935$\pm$0.002 &240&7.48\\
 & 7155.347&0.46$\pm$0.04& 27&3.1$\pm$2.7& 0.53$\pm$0.00 & 0.3378$\pm$0.0001 & 0.928$\pm$0.002 &278&12.69\\
 & 7159.352&0.38$\pm$0.04& 21&9.8$\pm$2.0& 0.52$\pm$0.00 & 0.3350$\pm$0.0001 & 0.920$\pm$0.002 &290&13.71\\
 & 7160.355&0.54$\pm$0.04& 40&3.2$\pm$2.1& 0.54$\pm$0.00 & 0.3389$\pm$0.0001 & 0.932$\pm$0.001 &366&17.33\\
 & 7161.358&0.42$\pm$0.04& 27&1.1$\pm$2.1& 0.51$\pm$0.00 & 0.3357$\pm$0.0001 & 0.924$\pm$0.001 &352&18.51\\
 & 7168.372&0.41$\pm$0.04& 27&-3.7$\pm$2.2& 0.53$\pm$0.00 & 0.3359$\pm$0.0001 & 0.922$\pm$0.002 &281&11.10\\
 & 7169.393&0.40$\pm$0.04& 40&-11.5$\pm$2.2& 0.53$\pm$0.00 & 0.3337$\pm$0.0001 & 0.924$\pm$0.002 &268&10.58\\
\hline
HH Leo (2017)& 7841.501&0.60$\pm$0.04& & & 0.53$\pm$0.00 & 0.3416$\pm$0.0001 & 0.941$\pm$0.002 &329&15.29\\
 & 7842.466&0.49$\pm$0.04& & & 0.52$\pm$0.00 & 0.3412$\pm$0.0001 & 0.937$\pm$0.002 &277&12.09\\
 & 7847.378&0.51$\pm$0.03& & & 0.53$\pm$0.00 & 0.3430$\pm$0.0001 & 0.941$\pm$0.001 &268&7.71\\
 & 7851.464&0.48$\pm$0.04& & & 0.53$\pm$0.00 & 0.3399$\pm$0.0001 & 0.939$\pm$0.002 &262&11.33\\
 & 7852.399&0.42$\pm$0.04& & & 0.52$\pm$0.00 & 0.3398$\pm$0.0001 & 0.947$\pm$0.002 &310&13.41\\
 & 7854.483&0.49$\pm$0.04& & & 0.54$\pm$0.00 & 0.3405$\pm$0.0001 & 0.939$\pm$0.002 &311&13.58\\
 & 7855.431&0.47$\pm$0.04& & & 0.51$\pm$0.00 & 0.3364$\pm$0.0001 & 0.925$\pm$0.002 &319&15.82\\
 & 7856.452&0.49$\pm$0.04& & & 0.52$\pm$0.00 & 0.3375$\pm$0.0001 & 0.927$\pm$0.002 &309&13.46\\
 & 7857.384&0.44$\pm$0.04& & & 0.52$\pm$0.00 & 0.3370$\pm$0.0001 & 0.927$\pm$0.002 &327&15.45\\
 & 7860.425&0.49$\pm$0.04& & & 0.52$\pm$0.00 & 0.3398$\pm$0.0001 & 0.930$\pm$0.002 &267&11.30\\
 & 7861.374&0.52$\pm$0.03& & & 0.51$\pm$0.00 & 0.3383$\pm$0.0001 & 0.929$\pm$0.001 &286&10.05\\
 & 7863.444&0.59$\pm$0.04& & & 0.53$\pm$0.00 & 0.3385$\pm$0.0001 & 0.930$\pm$0.002 &256&10.22\\
 & 7864.384&0.60$\pm$0.04& & & 0.53$\pm$0.00 & 0.3399$\pm$0.0001 & 0.936$\pm$0.002 &290&14.13\\
 & 7865.462&0.56$\pm$0.04& & & 0.55$\pm$0.00 & 0.3429$\pm$0.0001 & 0.940$\pm$0.002 &313&13.01\\
\hline  
\end{longtable}
\noindent
        \textbf{Notes:} Activity data, and longitudinal magnetic fields are from the respective sources, 1: \cite{folsom:2016}, 2: \cite{folsom:2018}. $*$ $\chi^2$ is calculated with the default errors, not the scaled errors from the MCMC-inference.        
}
\begin{table}[h]
    \centering
    \caption{BIC-values from the magnetic inference}
    \label{tab:BIC}
    \begin{tabular}{lrrrrr}
    \hline
    \hline
        \#-comp & HIP76768 & TYC 6349-0200-1 & Mel 25-5& HH Leo 2015 & 2017\\
        \hline
        2 & -618 & -1658 & -574  & -370 & -368\\ 
        3 & -650 & -1703 & -643 & -374 & -371\\ 
        4 & -708 & -1730 & -641 & -372 & -370\\ 
        5 & -777 & -1723 & -639 & -- & --\\ 
        6 & -811 & -- & -- & -- & --\\ 
        7 & -808 & -- & -- & -- & --\\ 
    \hline
    \end{tabular}
    \tablefoot{{The first column represents the number of magnetic field strengths (including B = 0) used in the model.} -- represents a model that was not used in the inference.}
\end{table}
\newpage
\section{Representative fit}
\label{app:repfit}
\begin{figure*}[h]
    \centering
    \includegraphics[width=0.8\linewidth]{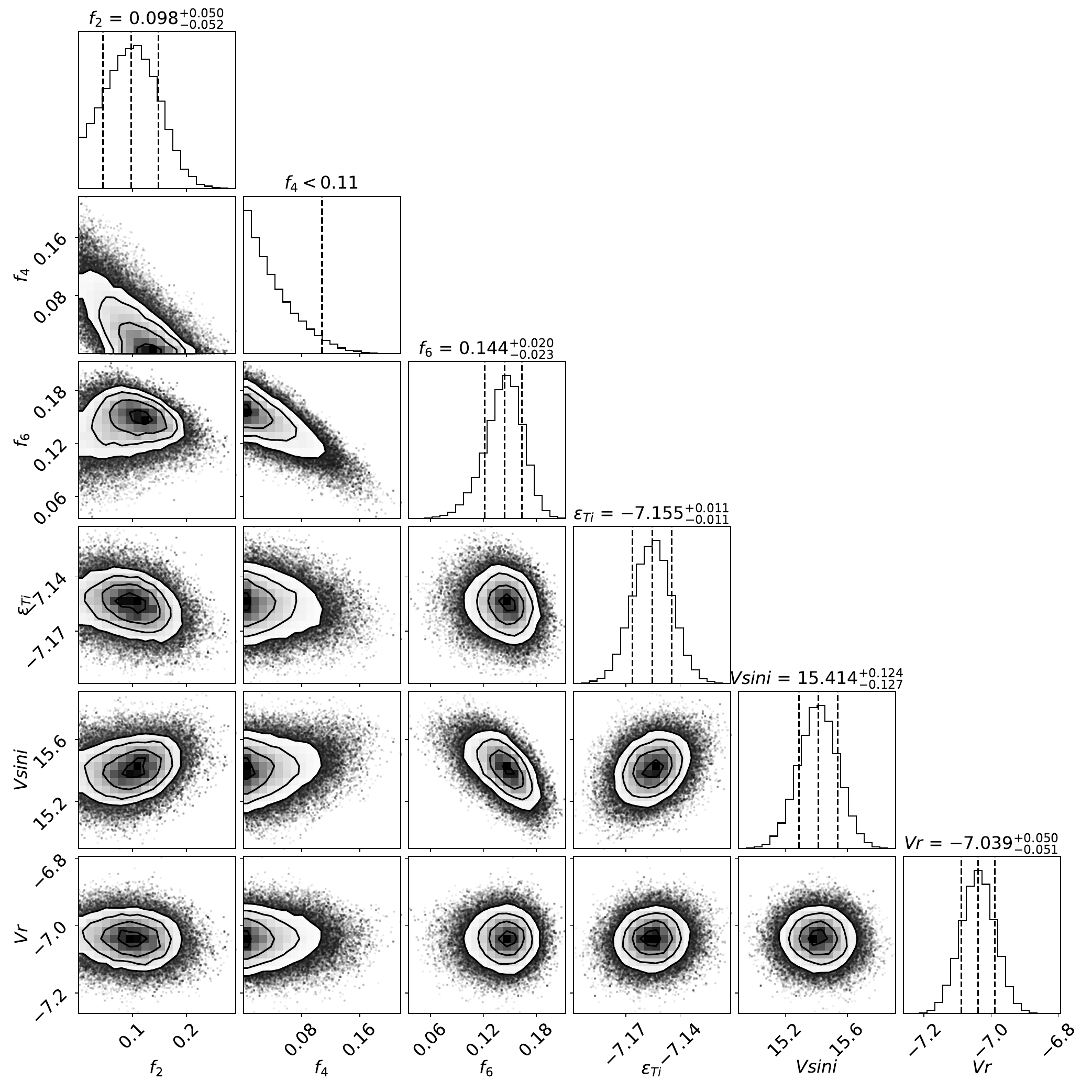}
    \caption{Corner plot from the MCMC inference of the mean spectrum of TYC 6349-0200-1. {The contours correspond to 0.5, 1, 1.5, 2\,$\sigma$ for a 2D Gaussian distribution.} Histogram in the second column shows 95\,\% upper limits of the $f_4$ parameter.}
    \label{fig:corner}
\end{figure*}
\begin{figure}[h]
    \centering
    \includegraphics[width=\linewidth]{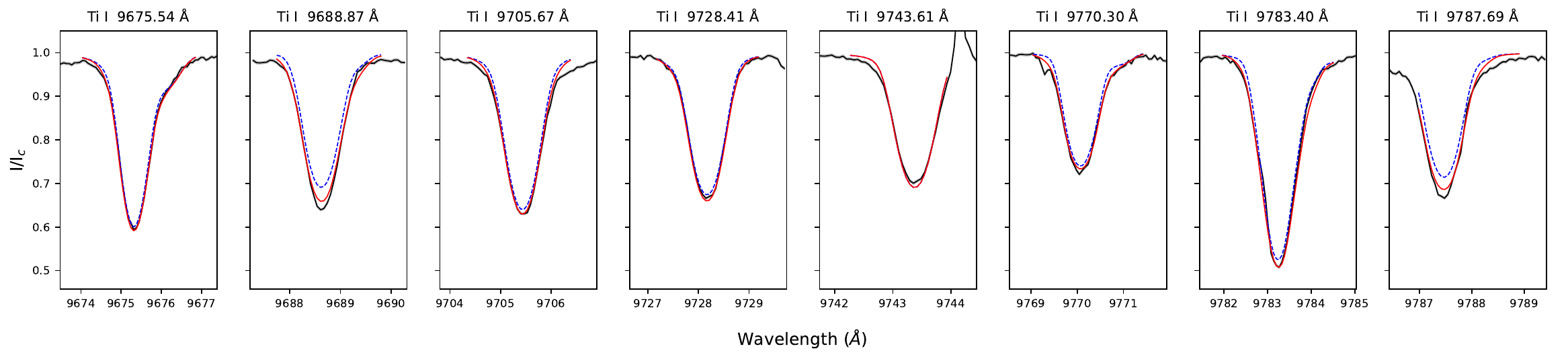}
    \caption{Best fit obtained from the MCMC inference of TYC 6349-0200-1.}
    \label{fig:spec}
\end{figure}
\end{appendix}
\end{document}